\documentclass[journal=jpcafh,layout=twocolumn,manuscript=letter]{achemso}
\setkeys{acs}{doi = true}
\newcommand{\doi}[1]{\href{http://dx.doi.org/#1}{\nolinkurl{#1}}}

\usepackage{amssymb}
\usepackage{amsmath}
\usepackage{amsfonts}
\usepackage{booktabs}
\usepackage{siunitx}
\usepackage[hidelinks]{hyperref}
\usepackage{mathrsfs}
\usepackage{mathtools}
\usepackage{multirow}
\usepackage{pdflscape}
\usepackage{physics}
\usepackage{subfig}

\usepackage{titlesec}

\usepackage[fontsize=11pt]{scrextend}
\titleformat{\section}
{\normalfont\sffamily\bfseries}
{\thesection.}{0.25em}{\uppercase}

\titleformat{\subsection}[runin]
{\normalfont\sffamily\bfseries}
{\thesubsection}{0.25em}{}[.\;\;]

\titleformat{\suppinfo}
{\normalfont\sffamily\bfseries}
{\thesubsection}{0.25em}{}

\titlespacing*{\section}{0pt}{0.5\baselineskip}{0.01\baselineskip}
\titlespacing*{\subsection}{0pt}{0.125\baselineskip}{0.01\baselineskip}

\newcommand{\spqe}[1]{SPQE}
\newcommand{\adaptsd}[1]{ADAPT-VQE-SD}
\newcommand{\adaptgsd}[1]{ADAPT-VQE-GSD}

\title{Unitary Coupled Cluster: Seizing the Quantum Moment}

\author{Ilias Magoulas}
\email{ilias.magoulas@emory.edu}
\affiliation
{Department of Chemistry and Cherry Emerson Center for Scientific Computation,
	Emory University, Atlanta, Georgia 30322, USA}

\author{Francesco A.\ Evangelista}
\email{francesco.evangelista@emory.edu}
\affiliation
{Department of Chemistry and Cherry Emerson Center for Scientific Computation,
	Emory University, Atlanta, Georgia 30322, USA}

\date{\today}

\begin{document}
	
	\begin{abstract}
		
Shallow, CNOT-efficient quantum circuits are crucial for performing accurate computational chemistry simulations
on current noisy quantum hardware. Here, we explore the usefulness of non-iterative energy corrections, based on
the method of moments of coupled-cluster theory, for accelerating convergence toward full configuration interaction.
Our preliminary numerical results relying on iteratively constructed ans\"{a}tze suggest that chemically accurate
energies can be obtained with substantially more compact circuits, implying enhanced resilience to gate and
decoherence noise.
		
	\end{abstract}
	
	\maketitle
	
	The first electronic structure calculation on quantum hardware was reported more than 10 years
	ago.\cite{Lanyon.2010.10.1038/NCHEM.483}
	It involved the iterative quantum phase estimation computation of the spectrum of $\text{H}_2$ in a
	minimum basis set. Since then, in tandem with various technological innovations, significant
	algorithmic advances have taken place aimed at the full utilization of the current generation of noisy
	intermediate-scale quantum devices.\cite{Bharti.2022.10.1103/RevModPhys.94.015004} Of particular
	importance is the arsenal of hybrid quantum--classical
	approaches,\cite{Endo.2021.10.7566/JPSJ.90.032001,Callison.2022.10.1103/PhysRevA.106.010101}
	including the variational (VQE),\cite{Peruzzo.2014.10.1038/ncomms5213,McClean.2016.10.1088/1367-2630/18/2/023023,
		Cerezo.2021.10.1038/s42254-021-00348-9,Tilly.2022.10.1016/j.physrep.2022.08.003,Fedorov.2022.10.1186/s41313-021-00032-6}
	contracted,\cite{Smart.2021.10.1103/PhysRevLett.126.070504} and
	projective\cite{Stair.2021.10.1103/PRXQuantum.2.030301} (PQE) quantum eigensolvers,
	quantum imaginary time evolution,\cite{Motta.2020.10.1038/s41567-019-0704-4,Sun.2021.10.1103/PRXQuantum.2.010317}
	and quantum subspace diagonalization techniques,\cite{Motta.2020.10.1038/s41567-019-0704-4,
		McClean.2017.10.1103/PhysRevA.95.042308,Parrish.2019.1909.08925v1,Stair.2020.10.1021/acs.jctc.9b01125,
		Huggins.2020.10.1088/1367-2630/ab867b} to mention a few.
	Such schemes require much shallower circuits than pure quantum algorithms, e.g., quantum phase
	estimation,\cite{Kitaev.1995.quant-ph/9511026,Abrams.1997.10.1103/PhysRevLett.79.2586,
		Abrams.1999.10.1103/PhysRevLett.83.5162} substantially reducing the computational cost
	and sensitivity to gate and decoherence noise.
	
	To pave the way toward practical and robust computations on actual quantum devices,
	various schemes have been devised to reduce the circuit depth of hybrid quantum--classical
	algorithms even further. In the case of ansatz-dependent approaches, which are of particular
	importance to this work, the construction of compact, yet sufficiently expressive, trial states
	is crucial. In general, the trial state is expressed as
	\begin{equation}\label{eq_trial_state}
		\ket*{\Psi^{(T)}(\mathbf{t})} = U^{(T)}(\mathbf{t}) \ket*{\Phi},
	\end{equation}
	where $\mathbf{t} = (t_1, t_2, \ldots)$ is a vector of parameters, $\ket*{\Phi}$ is a reference state
	that can be easily realized on the quantum device, and
	\begin{equation}\label{eq_unitary}
		U^{(T)} (\mathbf{t}) = \prod_\mu U_\mu^{(T)}(t_\mu)
	\end{equation}
	is the unitary operator that rotates $\ket*{\Phi}$ to the trial state $\ket*{\Psi^{(T)}(\mathbf{t})}$.
	Equation \eqref{eq_unitary} immediately reveals two complementary strategies for reducing the circuit
	complexity of ansatz-dependent algorithms. First, it is beneficial to minimize the number of
	parameters entering $U^{(T)}$, without, however, significantly sacrificing the expressivity of the ansatz. It is, thus,
	not surprising that fixed ans\"{a}tze have been almost invariably replaced by their iteratively
	constructed counterparts.\cite{Grimsley.2019.10.1038/s41467-019-10988-2,
		Tang.2021.10.1103/PRXQuantum.2.020310,Yordanov.2021.10.1038/s42005-021-00730-0,
		Ryabinkin.2020.10.1021/acs.jctc.9b01084,Stair.2021.10.1103/PRXQuantum.2.030301,
		Smart.2021.10.1103/PhysRevLett.126.070504,Mazziotti.2021.10.1088/1367-2630/ac3573,
		Smart.2022.10.1103/PhysRevA.105.062424,Fedorov.2022.10.22331/q-2022-05-02-703}
	In doing so, one eliminates the typically large number of superfluous parameters inadvertently
	incorporated in fixed-ans\"{a}tze approaches. Second, the form of the individual unitaries $U_\mu^{(T)}$ should
	be such that the corresponding quantum circuits are as compact as possible. 
	To that end, different flavors of ans\"{a}tze have been explored, typically derived from the unitary 
	extension\cite{Kutzelnigg.1977.10.1007/978-1-4757-0887-5_5,Kutzelnigg.1982.10.1063/1.444231,
		Kutzelnigg.1983.10.1063/1.446313,Kutzelnigg.1984.10.1063/1.446736,Bartlett.1989.10.1016/S0009-2614(89)87372-5,
		Szalay.1995.10.1063/1.469641,Taube.2006.10.1002/qua.21198,Cooper.2010.10.1063/1.3520564,
		Evangelista.2011.10.1063/1.3598471,Harsha.2018.10.1063/1.5011033,Filip.2020.10.1063/5.0026141,
		Freericks.2022.10.3390/sym14030494,Anand.2022.10.1039/d1cs00932j}
	of coupled-cluster theory\cite{Coester.1958.10.1016/0029-5582(58)90280-3,
		Coester.1960.10.1016/0029-5582(60)90140-1,Cizek.1966.10.1063/1.1727484,Cizek.1969.10.1002/9780470143599.ch2,
		Cizek.1971.10.1002/qua.560050402,Paldus.1972.10.1103/PhysRevA.5.50}
	(UCC). In this case, the $U_\mu^{(T)}(t_\mu)$ unitaries have the form
	\begin{equation}\label{eq_elementary_unitary}
		U_\mu^{(T)}(t_\mu) = e^{t_\mu \kappa_\mu},
	\end{equation}
	where $\kappa_\mu$ is a generic fermionic anti-Hermitian operator. In the language of
	second quantization and in the case of an $n$-body operator, $\kappa_\mu$ is defined as
	\begin{equation}
		\kappa_\mu \equiv 
		\kappa_{i_1 \ldots i_n}^{a_1 \ldots a_n} = a^{a_1} \cdots a^{a_n} a_{i_n} \cdots a_{i_1} - a^{i_1} 
		\cdots a^{i_n} a_{a_n} \cdots a_{a_1},
	\end{equation}
	where $a_p$ ($a^p \equiv a_p^\dagger$) is the annihilation (creation) operator
	acting on spinorbital $\phi_p$ and indices $i_1, i_2, \ldots$ or
	$i, j, \dots$ ($a_1, a_2, \ldots$ or $a, b, \ldots$) label spinorbitals occupied (unoccupied)
	in $\ket*{\Phi}$.
	Although starting from a fermionic UCC unitary and translating it to the qubit space is a
	natural choice for the study of electronic structure problems, it generally leads to circuits
	containing a substantial number of CNOT gates. This is particularly problematic, since current
	hardware implementations of two-qubit gates, such as CNOTs, are typically $\sim$10 times noisier than
	their one-qubit analogs.
	More CNOT-efficient ans\"{a}tze are obtained, for example, by directly building UCC in the qubit
	space\cite{Ryabinkin.2018.10.1021/acs.jctc.8b00932,Ryabinkin.2020.10.1021/acs.jctc.9b01084,
		Tang.2021.10.1103/PRXQuantum.2.020310} or
	employing the fermionic-excitation-based (FEB) circuits and their qubit (QEB)
	counterparts.\cite{Yordanov.2020.10.1103/PhysRevA.102.062612,Yordanov.2021.10.1038/s42005-021-00730-0,
		Xia.2021.10.1088/2058-9565/abbc74,Mazziotti.2021.10.1088/1367-2630/ac3573,
		Smart.2022.10.1103/PhysRevA.105.062424,Magoulas.2023.10.1021/acs.jctc.2c01016}
	Approximate implementations of the FEB and QEB circuits have recently been explored, resulting
	in single- and two-qubit gate counts that scale linearly with the excitation rank.\cite{Magoulas.2023.2304.12870}
	
	Another promising approach for reducing quantum resource requirements that has received relatively
	little attention, is the use of non-iterative energy corrections. Such schemes, in conjunction with
	iteratively constructed ans\"{a}tze, have the potential
	to accelerate convergence toward the exact, full configuration interaction (FCI), solution, thus
	decreasing the depth of the underlying quantum circuits. This avenue was initially explored in the
	context of the iterative qubit coupled cluster (iQCC)
	approach\cite{Ryabinkin.2020.10.1021/acs.jctc.9b01084} by Ryabinkin
	\latin{et al.}\cite{Ryabinkin.2021.10.1088/2058-9565/abda8e} By considering various
	low-order perturbative corrections to the iQCC energies, they were able to achieve a given level
	of accuracy with a smaller number of iQCC iterations.
	
	Similar in spirit to the work of Ryabinkin \latin{et al.}, in this study we explore the usefulness of
	non-iterative energy corrections resulting from the formalism of the method of moments of CC (MMCC)
	equations\cite{Piecuch.2000.10.1142/9789812792501_0001,Kowalski.2000.10.1063/1.481769,
		Kowalski.2000.10.1063/1.1290609,Piecuch.2001.10.1016/S0009-2614(01)00759-X,
		Fan.2005.10.1080/00268970500131595,Lodriguito.2006.10.1016/j.theochem.2006.03.014,
		Piecuch.2005.10.1063/1.2137318,Piecuch.2006.10.1016/j.cplett.2005.10.116,
		Wloch.2006.10.1080/00268970600659586,Wloch.2007.10.1021/jp072535l,
		Piecuch.2009.10.1002/qua.22367,Piecuch.2002.10.1080/0144235021000053811,
		Piecuch.2004.10.1007/s00214-004-0567-2,Shen.2012.10.1016/j.chemphys.2011.11.033}
	in accelerating the convergence to FCI of adaptive schemes. Corrections
	based on the MMCC framework have several desirable properties. Unlike
	non-iterative corrections based on many-body perturbation theory
	arguments,\cite{Urban.1985.10.1063/1.449067,
		Raghavachari.1985.10.1063/1.448718,Raghavachari.1989.10.1016/S0009-2614(89)87395-6,
		Kucharski.1989.10.1016/0009-2614(89)87388-9,Bartlett.1990.10.1016/0009-2614(90)87031-L,
		Kucharski.1993.10.1016/0009-2614(93)80186-S,Kucharski.1998.10.1063/1.475961,
		Kucharski.1998.10.1063/1.475962,Kucharski.1998.10.1063/1.476376,
		Musial.2000.10.1016/S0009-2614(00)00290-6}
	which generally fail when non-dynamical correlations become substantial, MMCC-type
	corrections are robust in the presence of quasi-degeneracies encountered in typical problems of chemical
	interest, such as single bond breaking and biradicals. In its exact form, the MMCC framework
	provides the non-iterative correction to the energy of an approximate CC method needed to recover
	the FCI energy. Although in practice one computes an approximate MMCC correction, improving its
	quality by incorporating higher-rank moments is straightforward, albeit computationally demanding
	on a classical machine. Nevertheless, in the case of a unitary ansatz, the necessary ingredients
	to compute such corrections can be efficiently calculated on a quantum computer (\latin{vide infra}).
	Furthermore, UCC-based approaches provide upper bounds to the FCI energy, rendering them
	immune to the catastrophic failures plaguing traditional
	single-reference CC methods in the presence of strong correlations, and excellent starting points
	for non-iterative corrections (see, also, ref \citenum{Fan.2005.10.1080/00268970500131595} for
	the generalization of the MMCC formalism to extended coupled cluster theory). 
	MMCC methods have been recently introduced to the realm of quantum computing, albeit from a
	different perspective. Peng and Kowalski have constructed a compact representation of
	non-unitary operators on quantum devices.\cite{Peng.2022.10.1103/PhysRevResearch.4.043172}
	This allowed them to devise quantum algorithms
	for computing MMCC corrections to conventional, \latin{i.e.}, non-unitary, CC schemes. Although
	Peng and Kowalski also outline how to extend the MMCC formalism to UCC ans\"{a}tze in a way similar
	to our approach, to the best of our knowledge the corresponding algorithm has never been implemented.

	We begin our discussion of non-iterative corrections by first summarizing the salient features of the
	quantum algorithms used to optimize the trial state.
	We employed the
	adaptive derivative-assembled pseudo-Trotterized VQE
	(ADAPT-VQE)\cite{Grimsley.2019.10.1038/s41467-019-10988-2} and selected PQE
	(SPQE)\cite{Stair.2021.10.1103/PRXQuantum.2.030301} approaches, which are based on 
	iteratively constructed ans\"{a}tze optimized with VQE and PQE, respectively.
	Recall that in the VQE and PQE schemes, the energy $E^{(T)}$ is computed as the expectation
	value of the Hamiltonian with respect to the trial state,
	\begin{equation}
		E^{(T)} = \ev*{H}{\Psi^{(T)}} = \ev*{\bar{H}^{(T)}}{\Phi},
	\end{equation}
	where $\bar{H}^{(T)} = {U^{(T)}}^\dagger H U^{(T)}$ is the similarity-transformed
	Hamiltonian.
	Thus, both VQE and PQE provide upper bounds to the FCI energy.
	At a high level,
	both algorithms rely on essentially the same two alternating steps. First, there is an expansion
	step promoting the
	growth of the ansatz. This is accomplished by ordering the operators in a
	given operator pool based on a predefined importance criterion and adding the most important
	operator(s) to the ansatz. In the case of VQE, the energy gradients
	\begin{equation}
		g_\mu (T) \equiv \frac{\partial E^{(T)}}{\partial t_\mu}
	\end{equation}
	play the role of the importance criterion, while in PQE the residuals of the UCC
	equations are used instead. Recall that the UCC residuals are projections onto the manifold of excited Slater determinants ($\ket{\Phi_\mu} \equiv \kappa_\mu \ket{\Phi}$) of the connected
	cluster form of the Schr\"{o}dinger equation with the UCC ansatz, eq \eqref{eq_trial_state},
	\begin{equation}
		r_\mu (T) \equiv \mel*{\Phi_\mu}{\bar{H}^{(T)}}{\Phi}.
	\end{equation}
	Note that, as shown in ref \citenum{Stair.2021.10.1103/PRXQuantum.2.030301},
	the residuals of all operators in the pool can be efficiently estimated
	by repeated measurements of the ``residual state,'' defined as
	\begin{equation}
		\ket*{\tilde{r}(T)} = {U^{(T)}}^\dagger e^{i \Delta t H} U^{(T)} \ket{\Phi}.
	\end{equation}
	The second step of the algorithm is the optimization of the ansatz parameters. 
	If we partition the full operator space into those incorporated in the ansatz ($P$) and those excluded ($Q$), then 
	for VQE the
	optimization is performed variationally, which translates into enforcing the condition
		$g_p = 0 \; \forall \kappa_p \in P$.
  	In the PQE case, a similar
	condition is imposed, this time, however, for the residuals ($r_p = 0 \; \forall \kappa_p \in P$). As shown in ref \citenum{Stair.2021.10.1103/PRXQuantum.2.030301}, the exact evaluation
	of a residual element in PQE has the same cost as the exact estimation of the corresponding gradient
	element in VQE using the parameter shift rule.\cite{Schuld.2019.10.1103/PhysRevA.99.032331,
		Kottmann.2021.10.1039/d0sc06627c}
	The remaining details of the SPQE and ADAPT-VQE algorithms can be found in refs
	\citenum{Stair.2021.10.1103/PRXQuantum.2.030301} and \citenum{Grimsley.2019.10.1038/s41467-019-10988-2},
	respectively.

	Next, we move on with the discussion of the non-iterative corrections. Here, we consider two families of
	non-iterative MMCC-type  corrections. The first one has the form
	\begin{equation}\label{eq_mmccIa}
		\delta_\text{Ia}(T) = \sum_{\mu} \frac{\abs{r_\mu(T)}^2}{E^{(T)} - E_\mu^{(T)}},
	\end{equation}
	where $r_\mu(T)$ are the residuals or moments of the UCC equations\cite{Jankowski.1991.10.1007/BF01117411,
		Piecuch.2000.10.1142/9789812792501_0001,Kowalski.2000.10.1063/1.481769}
	and $E_\mu^{(T)} = \ev*{\bar{H}^{(T)}}{\Phi_\mu}$ the diagonal elements of $\bar{H}^{(T)}$.
	Note that the summation over $\mu$ excludes the reference determinant.
	Equation \eqref{eq_mmccIa} can be regarded as the direct translation of  biorthogonal
	MMCC expansions\cite{Piecuch.2005.10.1063/1.2137318,Piecuch.2006.10.1016/j.cplett.2005.10.116,
		Wloch.2006.10.1080/00268970600659586} and their
	CC(\textit{P};\textit{Q}) generalization\cite{Shen.2012.10.1016/j.chemphys.2011.11.033,
		Shen.2012.10.1063/1.3700802,Shen.2012.10.1021/ct300762m} to the language of UCC. In the
	numerator of the CC(\textit{P};\textit{Q}) correction appears the product of moments of CC equations
	corresponding to the right and left CC states. Recall that traditional CC theory is non-Hermitian and as such
	the CC similarity-transformed Hamiltonian has distinct right and left eigenvectors.
	Since the UCC similarity-transformed
	Hamiltonian is Hermitian, in the translation process, the left-state moment is replaced by the
	complex conjugate of the right moment. From an alternative point of view,
	eq \eqref{eq_mmccIa} can be derived using L\"{o}wdin's partitioning procedure taking the zeroth-order part of
	the Hamiltonian to be the diagonal elements of $\bar{H}^{(T)}$.\cite{Stanton.1997.10.1016/S0009-2614(97)01144-5}

	In our current implementation, the summation appearing in eq \eqref{eq_mmccIa} is performed over
	all excited Slater determinants $\ket*{\Phi_\mu}$ for which $r_\mu (T) \ne 0$, namely,
	$\ket*{\Phi_\mu} \in Q$ for PQE and $\ket*{\Phi_\mu} \in P \oplus Q$ in the case of VQE. For example, in a
	UCCSD calculation relying on PQE optimization (UCCSD-PQE), the optimum parameters are obtained
	by imposing the residual conditions
	\begin{equation}
		r_a^i (\text{UCCSD-PQE}) \equiv \mel*{\Phi_i^a}{\bar{H}^{(\text{UCCSD-PQE})}}{\Phi} = 0
	\end{equation}
	and
	\begin{equation}
		r_{ab}^{ij} (\text{UCCSD-PQE}) \equiv \mel*{\Phi_{ij}^{ab}}{\bar{H}^{(\text{UCCSD-PQE})}}{\Phi} = 0.
	\end{equation}
	Thus, the summation in eq \eqref{eq_mmccIa} defining the $\delta_\text{Ia} (\text{UCCSD-PQE})$ moment correction
	runs over all triply, quadruply, \latin{etc.}\ excited Slater determinants (note that, unlike the
	case of traditional CCSD where the moment expansion naturally terminates after hextuples, for
	UCCSD-PQE all moments corresponding to higher-than-double excitations are, in principle, non-zero).
	Alternatively, in a VQE-based UCCSD (UCCSD-VQE) computation, the parameter optimization is performed
	variationally, \latin{i.e.}, by enforcing
	\begin{equation}
		g_a^i \equiv \frac{\partial E^\text{(UCCSD-VQE)}}{\partial t_a^i} = 0
	\end{equation}
	and
	\begin{equation}
		g_{ab}^{ij} \equiv \frac{\partial E^\text{(UCCSD-VQE)}}{\partial t_{ab}^{ij}} = 0.
	\end{equation}
	Consequently, since no residual condition is imposed, the $\delta_\text{Ia} (\text{UCCSD-VQE})$
	non-iterative correction involves all UCCSD-VQE moments, including those corresponding
	to singles and doubles. In this case, the appropriate schemes are based on the generalized MMCC
	formalism that was introduced in the context of extended
	coupled cluster theory.\cite{Fan.2005.10.1080/00268970500131595}
	
	Next, we consider the computational burden associated with the evaluation of eq \eqref{eq_mmccIa}. To
	this end, we examine the availability of its constituents. In the case of SPQE,
	the $r_\mu (\text{SPQE})$ residuals are already available, since they are generated for the selection of new operators.
	This is analogous to how perturbative energy corrections are evaluated in selected configuration interaction (CI) techniques, such as the CI method using perturbative selection made
	iteratively,\cite{Huron.1973.10.1063/1.1679199,Garniron.2017.10.1063/1.4992127,
		Garniron.2019.10.1021/acs.jctc.9b00176} adaptive CI,\cite{Schriber.2016.10.1063/1.4948308,
		Schriber.2017.10.1021/acs.jctc.7b00725} and adaptive sampling CI,\cite{Tubman.2016.10.1063/1.4955109}
	to mention a few. In the case of ADAPT-VQE simulations, the residuals need to be evaluated, but this can be
	efficiently accomplished by repeated measurements
	of the residual vector, as discussed above. Furthermore, both SPQE and ADAPT-VQE have access to the
	$E^{(T)}$ energy of the trial state, as part of the parameter optimization process. Thus, the major
	computational overhead is the evaluation
	of the diagonal elements of the similarity-transformed Hamiltonian in every macro-iteration.
	To reduce the computational cost associated with the evaluation of $E^{(T)}_\mu$, we also consider two approximations to the denominators that enter into eq \eqref{eq_mmccIa}.
	In the first one, denoted as $\delta_\text{Ib} (T)$,
	we replace the similarity-transformed Hamiltonian in $E_\mu^{(T)}$ by the bare Hamiltonian,
	\begin{equation}\label{eq_mmccIb}
		\delta_\text{Ib}(T) = \sum_\mu \frac{\abs{r_\mu(T)}^2}{E^{(T)} - E_\mu}.
	\end{equation}
	This allows us to efficiently evaluate the $E_\mu = \ev*{H}{\Phi_\mu}$ energies only once per simulation using a classical algorithm. The second
	approximation, designated as $\delta_\text{Ic} (T)$, is more drastic, replacing the entire denominator
	appearing in eq \eqref{eq_mmccIa} by its M{\o}ller--Plesset counterpart,
	\begin{equation}\label{eq_mmccIc}
		\delta_\text{Ic}(T) = \sum_\mu \frac{\abs{r_\mu(T)}^2}{\Delta_\mu},
	\end{equation}
	where $\Delta_\mu \equiv \Delta_{i_1 \ldots i_n}^{a_1 \ldots a_n} = \epsilon_{i_1} + \cdots
	+ \epsilon_{i_n} - \epsilon_{a_1} - \epsilon_{a_n}$ with $\epsilon_p$ denoting the Hartree--Fock
	energy of spinorbital $\phi_p$. Note that unlike eqs \eqref{eq_mmccIa} and \eqref{eq_mmccIc}, the
	approximation defined in eq \eqref{eq_mmccIb} is not rigorously size consistent [see the Supporting Information
	for the proof and underlying conditions for the size consistency of eqs \eqref{eq_mmccIa} and \eqref{eq_mmccIc}].
	Nevertheless, our preliminary numerical results suggest that this is not a major issue (\latin{vide infra}).
	 
	The second family of non-iterative corrections we consider is based on the one defining the original renormalized
	and completely renormalized CC approaches,\cite{Piecuch.2000.10.1142/9789812792501_0001,
		Kowalski.2000.10.1063/1.481769,Kowalski.2000.10.1063/1.1290609} namely,
	\begin{equation}\label{eq_mmccII}
		\delta_\text{II} (T) = \sum_\mu \frac{\mel*{\Psi^\text{(ext)}}{U^{(T)}}{\Phi_\mu} r_\mu (T)}
		{\braket{\Psi^\text{(ext)}}{\Psi^{(T)}}}.
	\end{equation}
	In eq \eqref{eq_mmccII}, $\ket*{\Psi^\text{(ext)}}$ is the wavefunction from an external source and serves
	as an approximation to the FCI wavefunction. If $\ket*{\Psi^\text{(ext)}} \equiv
	\ket*{\Psi^\text{(FCI)}}$, then eq \eqref{eq_mmccII} yields the exact MMCC correction, \latin{i.e.},
	$E^{(T)} + \delta_\text{II}(T) = E^\text{(FCI)}$. In this regard, eq \eqref{eq_mmccII} shares the same
	philosophy with externally corrected CC approaches.\cite{Paldus.1984.10.1103/PhysRevA.30.2193,
		Piecuch.1990.10.1007/BF01119191,Piecuch.1996.10.1103/PhysRevA.54.1210,Paldus.1994.10.1007/BF01123868,
		Stolarczyk.1994.10.1016/0009-2614(93)E1333-C,Peris.1997.10.1002/(SICI)1097-461X(1997)62:2<137::AID-QUA2>3.0.CO;2-X,
		Peris.1999.10.1063/1.479116,Li.1997.10.1063/1.474289,Li.2006.10.1063/1.2194543,
		Paldus.2017.10.1007/s10910-016-0688-6,Deustua.2018.10.1063/1.5055769,
		Aroeira.2020.10.1021/acs.jctc.0c00888,Magoulas.2021.10.1021/acs.jctc.1c00181}
	The derivation of this non-iterative correction can be found in the Appendix of Peng and
	Kowalski.\cite{Peng.2022.10.1103/PhysRevResearch.4.043172}
	As was the case with eqs \eqref{eq_mmccIa}, \eqref{eq_mmccIb}, and \eqref{eq_mmccIc}, the summation appearing in eq \eqref{eq_mmccII} involves all
	excited Slater determinants $\ket*{\Phi_\mu}$ for which $r_\mu (T) \ne 0$.
	An intriguing aspect of this non-iterative correction is its flexibility regarding the external state. One has
	the possibility of utilizing any quantum algorithm that can produce an approximate eigenstate of the Hamiltonian
	and combining it with any ansatz-dependent scheme with the intention of improving the results of both.

	To assess the ability of the non-iterative, MMCC-type, energy corrections explored in this study to accelerate convergence toward FCI and reduce the resources used by the underlying quantum algorithms,
	we performed numerical simulations of the symmetric dissociation of the
	$\text{H}_6$ linear chain, as described by the minimum STO-6G basis.\cite{Hehre.1969.10.1063/1.1672392}
	In these preliminary calculations, we considered three representative points along the potential energy
	curve, characterized by the distances between neighboring H atoms of $R_\text{H--H} = 1.0$,
2.0, and \SI{3.0}{\AA}, corresponding to the equilibrium, recoupling, and strongly correlated
	regions of the potential, respectively. The underlying quantum algorithms used to gauge the performance
	of the $\delta_\text{I} (T)$ corrections, eqs \eqref{eq_mmccIa}, \eqref{eq_mmccIb}, and \eqref{eq_mmccIc},
	were SPQE and ADAPT-VQE. To further reduce the depth of the resulting
	circuits, we employed the CNOT-efficient FEB-SPQE (fermionic operators) and QEB-ADAPT-VQE (qubit operators) variants. In the case of SPQE,
	we also considered the approximate FEB scheme (aFEB), which has been shown to faithfully reproduce the
	parent FEB-SPQE energetics with negligible symmetry breaking and a linear CNOT count.\cite{Magoulas.2023.2304.12870}
	In benchmarking the $\delta_\text{II} (T)$ correction, the trial state was obtained with VQE using a UCCSD ansatz. As a proof of principle,
	the external states were initially taken to be UCCGSD and UCCSDTQPH, optimized with VQE and PQE,
	respectively. Both UCCGSD and UCCSDTQPH are exact in the case of $\text{H}_6$/STO-6G and the $\delta_\text{II} (\text{UCCSD})$ correction to the UCCSD energies reproduced the FCI result
	with microhartree or better accuracy. In the production run, the external states were those
	extracted at every macro-iteration of QEB-ADAPT-VQE simulations. This approach shares the same
	philosophy with the externally corrected CCSD methods based on FCI quantum Monte
	Carlo\cite{Deustua.2018.10.1063/1.5055769} and
	selected CI wavefunctions\cite{Aroeira.2020.10.1021/acs.jctc.0c00888,Magoulas.2021.10.1021/acs.jctc.1c00181}
	(see, also, the semi-stochastic  CC(\textit{P};\textit{Q}) scheme based on FCI quantum Monte
	Carlo\cite{Deustua.2017.10.1103/PhysRevLett.119.223003}
	and its selected CI counterpart\cite{Gururangan.2021.10.1063/5.0064400}).
	
	All SPQE simulations employed a full operator pool and macro-iteration threshold of \SI{e-2}{\textit{E}_h}.
	The PQE micro-iteration threshold was set to \SI{e-5}{\textit{E}_h} while the direct inversion of
	the iterative subspace\cite{Pulay.1980.10.1016/0009-2614(80)80396-4,Pulay.1982.10.1002/jcc.540030413,
		Scuseria.1986.10.1016/0009-2614(86)80461-4} was utilized to accelerate convergence. Although in
	typical SPQE runs, multiple operators are added to the ansatz per macro-iteration, here we followed
	the ADAPT-VQE paradigm and added operators sequentially. This allowed us to properly examine the
	ability of the non-iterative corrections to accelerate convergence of SPQE toward FCI. The ADAPT-VQE
	simulations relied on a pool of generalized singles and
	doubles\cite{Nooijen.2000.10.1103/PhysRevLett.84.2108,Nakatsuji.2000.10.1063/1.1287275}
	and macro- and micro-iteration thresholds of \SI{e-3}{\textit{E}_h} and \SI{e-5}{\textit{E}_h},
	respectively. The non-iterative energy corrections explored in this study have been implemented
	in a local version of QForte.\cite{Stair.2022.10.1021/acs.jctc.1c01155} All computations employed
	restricted Hartree--Fock references, with the one-
	and two-electron integrals obtained from Psi4.\cite{Smith.2020.10.1063/5.0006002}

	\begin{figure*}[!h]
	\centering
	\includegraphics[width=\textwidth]{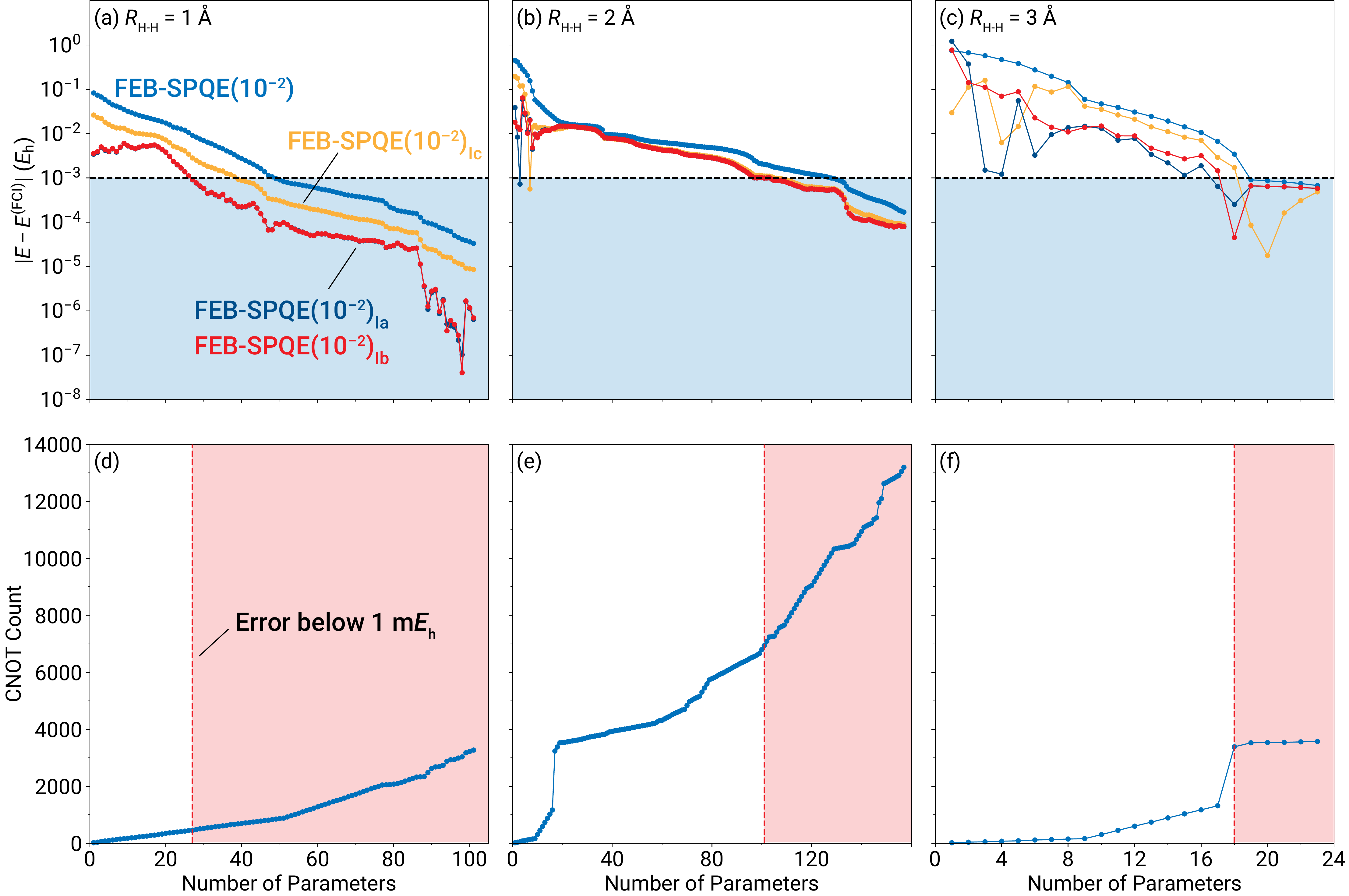}
	\caption{\label{figure1}
		Errors relative to FCI [(a)--(c)] and CNOT gate counts [(d)--(f)] characterizing the FEB-SPQE
		simulations of the symmetric dissociation of the $\text{H}_6$/STO-6G linear chain at three
		representative distances between neighboring H atoms, including $R_\text{H--H} = \SI{1.0}{\AA}$
		[(a) and (d)], $R_\text{H--H} = \SI{2.0}{\AA}$ [(b) and (e)], and $R_\text{H--H} = \SI{3.0}{\AA}$
		[(c) and (f)]. The blue-shaded area in the top-row panels indicates results within
		chemical accuracy (\SI{1}{\milli \textit{E}_h}) from FCI. The red-shaded area in the
		bottom-row panels denotes the CNOT counts of the underlying FEB-SPQE quantum circuits for which
		the FEB-SPQE$_\text{Ib}$ energies are within chemical accuracy. 
	}
	\end{figure*}
	We begin the discussion of our numerical results by examining the ability of the non-iterative,
	MMCC-type, energy corrections defined in eqs \eqref{eq_mmccIa}, \eqref{eq_mmccIb}, and \eqref{eq_mmccIc}
	to accelerate the convergence of FEB-SPQE simulations toward
	FCI. To that end, in the top panels to Figure \ref{figure1} we compare the energies resulting from
	FEB-SPQE and FEB-SPQE$_{\text{I}x}$ $\equiv$ FEB-SPQE+$\delta_{\text{I}x}$, with $x$ = a, b, c.
	 A quick
	inspection of Figure \ref{figure1} immediately reveals that all three corrections are capable of
	reproducing the FCI data within the chemical accuracy of \SI{1}{\milli \textit{E}_h} more rapidly
	than the underlying FEB-SPQE approach. Their performance is particularly impressive in the weakly
	and moderately correlated regions, requiring as much as $\sim$45\% and $\sim$20\% fewer parameters
	than FEB-SPQE. This emphasizes the exceptional ability of such non-iterative corrections to recover
	dynamical correlation effects. The decrease in the number of parameters is accompanied by similarly
	remarkable reductions in the CNOT counts of the corresponding ansatz circuits, as illustrated in the
	bottom panels to Figure \ref{figure1}. In comparing the various non-iterative
	corrections among themselves, the FEB-SPQE$_\text{Ic}$ scheme that relies on the standard
	M{\o}ller--Plesset denominator has the least favorable performance. With the exception of the very early stages
	of the FEB-SPQE simulations, FEB-SPQE$_\text{Ib}$ faithfully reproduces the results of its FEB-SPQE$_\text{Ia}$
	parent. This remains true even near the dissociation threshold of $\text{H}_6$, represented by the geometry
	characterized by $R_\text{H--H} = \SI{3.0}{\AA}$. This implies that the size inconsistency introduced in the denominator
	of eq \eqref{eq_mmccIb} is not severe, at least in the case of $\text{H}_6$. As a result, in the subsequent
	benchmarks we will be focusing on the $\delta_\text{Ib}$
	correction, since it not only provides results similar to the more complete $\delta_\text{Ia}$ scheme, but is also
	more computationally efficient (\latin{vide supra}).

	In an effort to reduce the circuit depth even further, we combine the $\delta_\text{Ib}$ non-iterative
	correction with aFEB-SPQE. As already mentioned above, the recently introduced aFEB circuits are
	approximate implementations of the CNOT-efficient FEB ones that require a number of CNOT gates
	that scales linearly with the excitation rank of a given operator.\cite{Magoulas.2023.2304.12870}
	Furthermore, although the aFEB approximation breaks the particle number $N$ and total spin
	projection $S_z$ symmetries, it has been demonstrated that aFEB-SPQE is characterized by an
	essentially negligible symmetry contamination and a faithful reproduction of the parent FEB-SPQE
	energetics. As might have been anticipated, the same is true when we examine the energies resulting
	from their $\delta_\text{Ib}$-corrected counterparts. Indeed,
	as shown in Figure S1 of the Supporting Information, the aFEB-SPQE$_\text{Ib}$ data is practically
	indistinguishable from those obtained with FEB-SPQE$_\text{Ib}$. This is even true in the recoupling
	region, characterized by $R_\text{H--H} = \SI{2.0}{\AA}$, where the largest deviation, on the order of
	\SI{0.1}{\milli \textit{E}_h}, is observed.	At this point, it is worth
	mentioning that due to the symmetry breaking introduced by the aFEB quantum circuits, aFEB-SPQE has non-zero
	moments corresponding to Slater determinants with $N \ne 6$ and/or $S_z \ne 0$ for the ground electronic state
	of $\text{H}_6$. In principle,
	one could incorporate such Fock-space moments into the MMCC-type non-iterative corrections to improve the quality
	of the results even further. However, since the contribution of the symmetry contaminants to the aFEB-SPQE
	wavefunction is negligible,\cite{Magoulas.2023.2304.12870} we did not consider such a generalized correction.
	
	\begin{figure*}[!h]
	\centering
	\includegraphics[width=\textwidth]{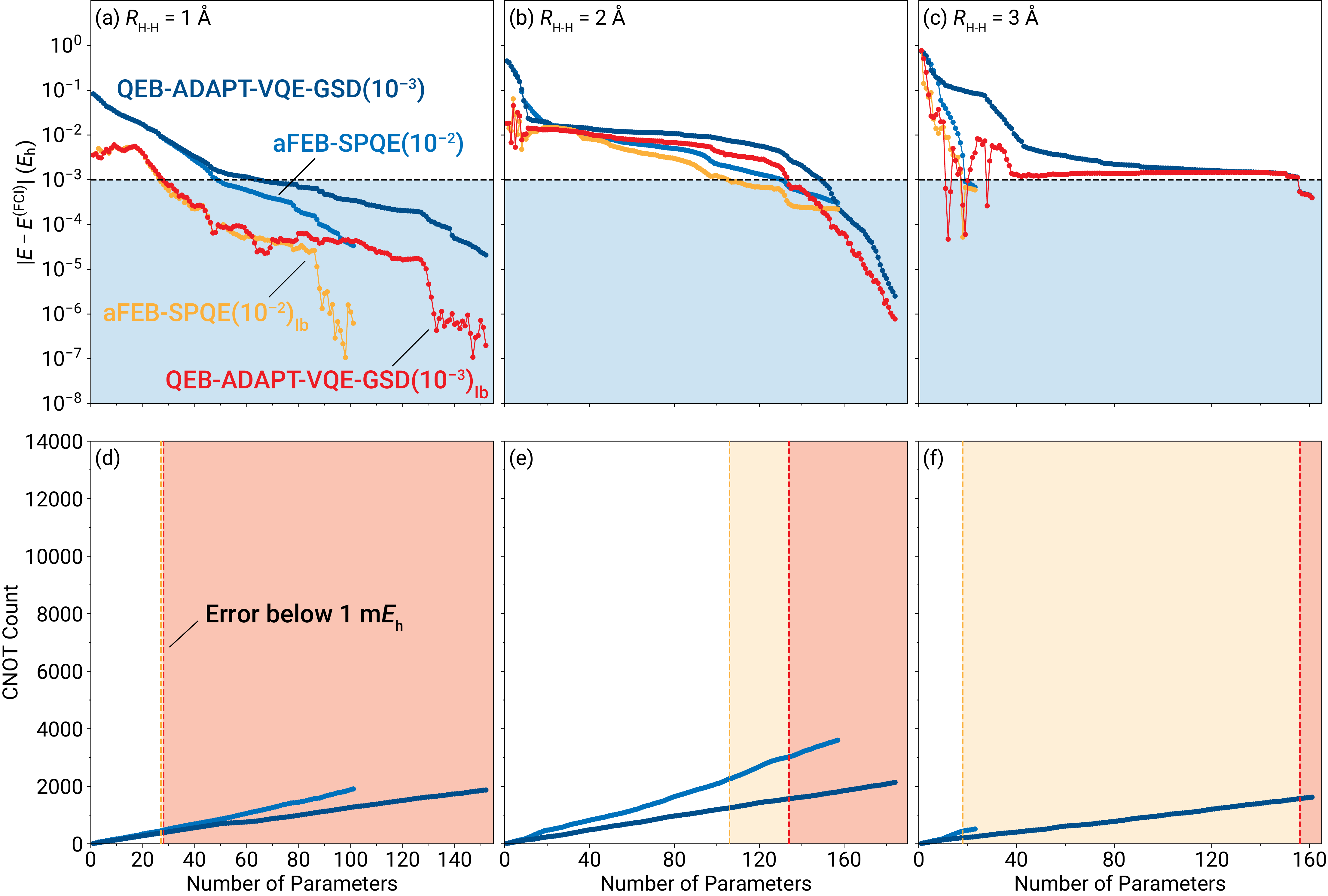}
	\caption{\label{figure2}
		Errors relative to FCI [(a)--(c)] and CNOT gate counts [(d)--(f)] characterizing the
		aFEB-SPQE and QEB-ADAPT-VQE-GSD
		simulations of the symmetric dissociation of the $\text{H}_6$/STO-6G linear chain at three
		representative distances between neighboring H atoms, including $R_\text{H--H} = \SI{1.0}{\AA}$
		[(a) and (d)], $R_\text{H--H} = \SI{2.0}{\AA}$ [(b) and (e)], and $R_\text{H--H} = \SI{3.0}{\AA}$
		[(c) and (f)]. The blue-shaded area in the top-row panels indicates results within
		chemical accuracy (\SI{1}{\milli \textit{E}_h}) from FCI. The yellow- and red-shaded areas in the
		bottom-row panels denotes the CNOT counts of the underlying aFEB-SPQE and QEB-ADAPT-VQE-GSD quantum circuits,
		respectively,  for which
		the aFEB-SPQE$_\text{Ib}$ and QEB-ADAPT-VQE-GSD$_\text{Ib}$ energetics are within chemical accuracy.
	}
	\end{figure*}
	Next, we examine the usefulness of the $\delta_\text{Ib}$ non-iterative correction in the context of
	ADAPT-VQE (QEB-ADAPT-VQE$_\text{Ib}$) and compare the results with those of aFEB-SPQE$_\text{Ib}$.
	When examining the errors with
	respect to FCI, we observe that, in general, the QEB-ADAPT-VQE$_\text{Ib}$ approach offers a major
	improvement over the underlying QEB-ADAPT-VQE energetics (top panels to Figure \ref{figure2}).
	This enables QEB-ADAPT-VQE$_\text{Ib}$ to
	recover the FCI data within \SI{1.0}{\milli \textit{E}_h} faster than the underlying QEB-ADAPT-VQE
	approach, resulting in substantial savings in terms of CNOTs (see bottom panels to Figure \ref{figure2}).
	Exception to this is the QEB-ADAPT-VQE simulation
	for the strongly correlated regime of $\text{H}_6$. In this case, QEB-ADAPT-VQE is characterized by a
	very slow convergence to FCI, most likely due to the presence of a local energy minimum. For
	QEB-ADAPT-VQE$_\text{Ib}$, this manifests itself as a plateau in the corresponding energies. The
	fact that after about 120 parameters the QEB-ADAPT-VQE and QEB-ADAPT-VQE$_\text{Ib}$ results practically coincide,
	\latin{i.e.}, the QEB-ADAPT-VQE moments are close to zero, further corroborates to the fact that the
	QEB-ADAPT-VQE wavefunction closely resembles that of an exact excited electronic state.
	Eventually both QEB-ADAPT-VQE and QEB-ADAPT-VQE$_\text{Ib}$ achieve chemical accuracy at the same time,
	requiring about 80\% of the number of parameters of FCI.

	Moving on to the comparison of the ADAPT-VQE and SPQE approaches, a quick inspection of the top
	panels to Figure \ref{figure2} immediately reveals that aFEB-SPQE provides, in general, more
	accurate energetics
	than QEB-ADAPT-VQE for the same number of parameters. Similar behavior is observed when we examine
	their MMCC-corrected counterparts. The fact that aFEB-SPQE$_\text{Ib}$ is characterized by a faster convergence
	to FCI than QEB-ADAPT-VQE$_\text{Ib}$ does not necessarily guarantee that it does so with fewer number of CNOTs.
	As demonstrated in the bottom panels to Figure \ref{figure2}, although both aFEB-SPQE and QEB-ADAPT-VQE
	generate quantum circuits with CNOT counts that scale, more or less, linearly with the number of parameters,
	the prefactor is smaller in the case of QEB-ADAPT-VQE. Indeed, for the three examined geometries
	of $\text{H}_6$, characterized by the $R_\text{H--H}$ values of \SI{1.0}{\AA}, \SI{2.0}{\AA}, and
	\SI{3.0}{\AA}, QEB-ADAPT-VQE$_\text{Ib}$ attains chemical accuracy with the same, less, and more CNOTs, respectively,
	than when aFEB-SPQE$_\text{Ib}$ does.

	\begin{figure*}[!h]
	\centering
	\includegraphics[width=\textwidth]{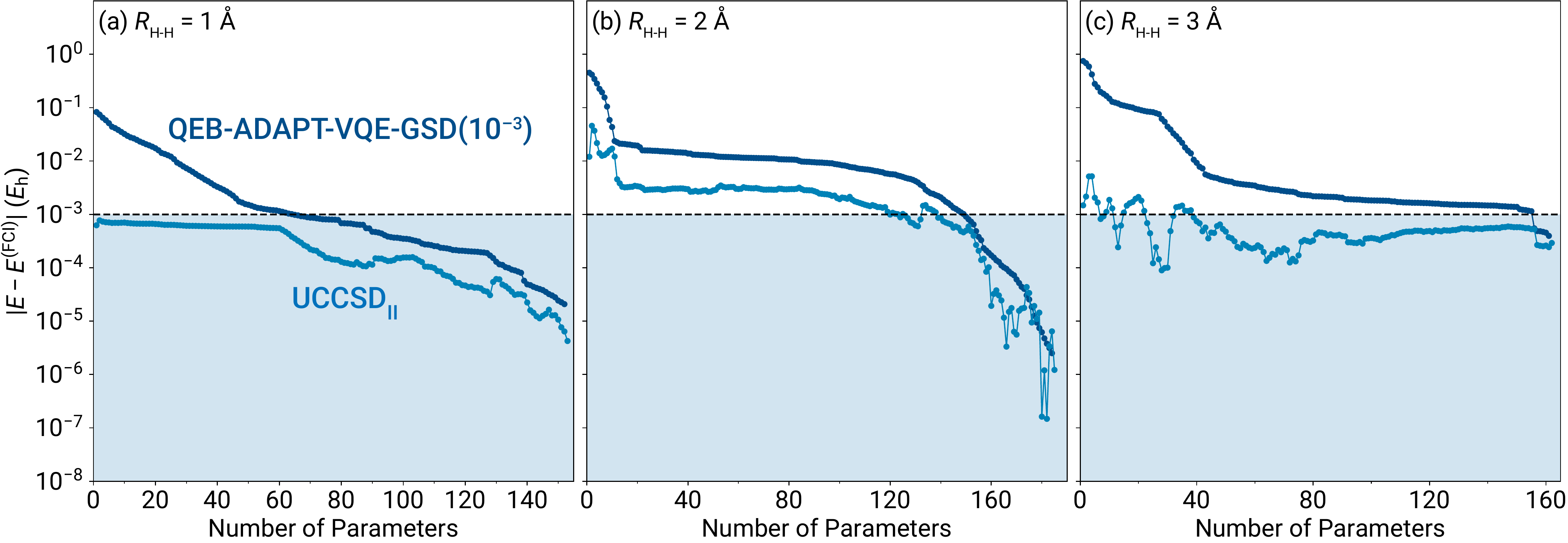}
	\caption{\label{figure3}
		Errors relative to FCI characterizing the moment-corrected UCCSD$_\text{II}$ scheme and the underlying
		QEB-ADAPT-VQE-GSD
		simulations of the symmetric dissociation of the $\text{H}_6$/STO-6G linear chain at three
		representative distances between neighboring H atoms, including (a) $R_\text{H--H} = \SI{1.0}{\AA}$,
		(b) $R_\text{H--H} = \SI{2.0}{\AA}$, and (c) $R_\text{H--H} = \SI{3.0}{\AA}$. The blue-shaded area
		indicates results within chemical accuracy (\SI{1}{\milli \textit{E}_h}) from FCI.
	}
	\end{figure*}
	Now we turn our attention to the $\delta_\text{II}$ MMCC-type correction, given by eq \eqref{eq_mmccII}. To gauge
	the performance of this correction, we considered the UCCSD$_\text{II}$ scheme in which we corrected the VQE UCCSD
	energies using the wavefunctions
	resulting from each macro-iteration defining the QEB-ADAPT-VQE simulations. The reason we opted to use
	QEB-ADAPT-VQE as the external source of the wavefunction rather than aFEB-SPQE is that, for the same number
	of parameters, the former generates the most CNOT-efficient circuits. In Figure \ref{figure3}, we present
	the errors relative to FCI characterizing the UCCSD$_\text{II}$ method and the underlying QEB-ADAPT-VQE approach.
	As depicted in Figure \ref{figure3}, with the exception of the very early stages of the QEB-ADAPT-VQE
	simulations, in which UCCSD is more accurate, UCCSD$_\text{II}$ consistently outperforms both UCCSD and
	QEB-ADAPT-VQE. As far as the number of CNOTs needed to achieve chemical accuracy is concerned, which of the UCCSD
	and QEB-ADAPT-VQE requires the deepest circuit depends on the strength of the correlation effects. Based on
	our numerical results for $\text{H}_6$, in the weakly and strongly correlated regions, where UCCSD$_\text{II}$
	recovers the FCI energies within \SI{1.0}{\milli \textit{E}_h} rather quickly, UCCSD is
	the bottleneck in terms of CNOT count. The situation changes when we examine the recoupling region, in which
	case QEB-ADAPT-VQE defines the overall CNOT cost.
	In comparing the UCCSD$_\text{II}$ and QEB-ADAPT-VQE$_\text{Ib}$ approaches among themselves,
	we observe that UCCSD$_\text{II}$ is characterized by a more rapid convergence to FCI. This can be mostly
	attributed to the fact that UCCSD offers a good starting point for the MMCC-type correction, especially so
	during the early stages of the simulation.

	In this work, we explored the usefulness of moment expansions, used in the past to define the
	renormalized and completely renormalized CC approaches, to construct non-iterative corrections
	to the energies obtained with hybrid quantum--classical algorithms employing a UCC ansatz.
	In an attempt to minimize the circuit depth, we combined the non-iterative energy corrections
	with iteratively constructed ans\"{a}tze and CNOT-efficient implementations of quantum circuits.
	Our preliminary numerical results for three representative geometries along the potential
	energy curve characterizing the symmetric dissociation of the $\text{H}_6$ linear chain
	are very encouraging. All of the examined corrections accelerated the convergence to FCI
	of the underlying quantum algorithms, resulting in substantial savings in terms of CNOT
	gates. The performance of the aFEB-SPQE$_\text{Ib}$, QEB-ADAPT-VQE$_\text{Ib}$, and
	UCCSD$_\text{II}$ is particularly promising.
	We point out that the corrections studied in this work are applicable even when the trial state is not of the UCC form,
	opening the possibility to improve hardware efficient ans\"{a}tze and other trial states that are not expressible via the UCC form.
	In the future, we will explore the flexibility of the
	$\delta_\text{II}$ correction, combing different types of trial and external wavefunctions.
	Finally, inspired by the success of MMCC-type corrections for excited electronic
	states,\cite{Kowalski.2004.10.1063/1.1632474,Wloch.2006.10.1080/00268970600659586,Piecuch.2009.10.1002/qua.22367,
		Lutz.2014.10.1016/j.comptc.2014.05.008,Piecuch.2015.10.1080/00268976.2015.1076901,Wloch.2005.10.1063/1.1924596,
		Fradelos.2011.10.1021/ct200101x,Shen.2012.10.1016/j.chemphys.2011.11.033,Yuwono.2020.10.1080/00268976.2020.1817592}
	we plan to investigate this aspect as well.
	

	\begin{acknowledgement}
		
		This work is supported by the U.S.\ Department of Energy under Award No.\ DE-SC0019374.
		
	\end{acknowledgement}

	\begin{suppinfo}
	
	\noindent Proof of the size consistency of the non-iterative corrections and additional numerical results.
	
	\noindent Numerical data generated in this study.
		
	\end{suppinfo}

	
	\bibliography{../../../Bibliography/refs}

\providecommand{\latin}[1]{#1}
\makeatletter
\providecommand{\doi}
  {\begingroup\let\do\@makeother\dospecials
  \catcode`\{=1 \catcode`\}=2 \doi@aux}
\providecommand{\doi@aux}[1]{\endgroup\texttt{#1}}
\makeatother
\providecommand*\mcitethebibliography{\thebibliography}
\csname @ifundefined\endcsname{endmcitethebibliography}
  {\let\endmcitethebibliography\endthebibliography}{}
\begin{mcitethebibliography}{119}
\providecommand*\natexlab[1]{#1}
\providecommand*\mciteSetBstSublistMode[1]{}
\providecommand*\mciteSetBstMaxWidthForm[2]{}
\providecommand*\mciteBstWouldAddEndPuncttrue
  {\def\EndOfBibitem{\unskip.}}
\providecommand*\mciteBstWouldAddEndPunctfalse
  {\let\EndOfBibitem\relax}
\providecommand*\mciteSetBstMidEndSepPunct[3]{}
\providecommand*\mciteSetBstSublistLabelBeginEnd[3]{}
\providecommand*\EndOfBibitem{}
\mciteSetBstSublistMode{f}
\mciteSetBstMaxWidthForm{subitem}{(\alph{mcitesubitemcount})}
\mciteSetBstSublistLabelBeginEnd
  {\mcitemaxwidthsubitemform\space}
  {\relax}
  {\relax}

\bibitem[Lanyon \latin{et~al.}(2010)Lanyon, Whitfield, Gillet, Goggin, Almeida,
  Kassal, Biamonte, Mohseni, Powell, Barbieri, Aspuru-Guzik, and
  White]{Lanyon.2010.10.1038/NCHEM.483}
Lanyon,~B.~P.; Whitfield,~J.~D.; Gillet,~G.~G.; Goggin,~M.~E.; Almeida,~M.~P.;
  Kassal,~I.; Biamonte,~J.~D.; Mohseni,~M.; Powell,~B.~J.; Barbieri,~M.
  \latin{et~al.}  Towards Quantum Chemistry on a Quantum Computer. \emph{Nat.
  Chem.} \textbf{2010}, \emph{2}, 106--111, DOI: \doi{10.1038/NCHEM.483}\relax
\mciteBstWouldAddEndPuncttrue
\mciteSetBstMidEndSepPunct{\mcitedefaultmidpunct}
{\mcitedefaultendpunct}{\mcitedefaultseppunct}\relax
\EndOfBibitem
\bibitem[Bharti \latin{et~al.}(2022)Bharti, Cerveta-Lierta, Kyaw, Haug,
  Alperin-Lea, Anand, Degroote, Heimonen, Kottmann, Menke, Mok, Sim, Kwek, and
  Aspuru-Guzik]{Bharti.2022.10.1103/RevModPhys.94.015004}
Bharti,~K.; Cerveta-Lierta,~A.; Kyaw,~T.~H.; Haug,~T.; Alperin-Lea,~S.;
  Anand,~A.; Degroote,~M.; Heimonen,~H.; Kottmann,~J.~S.; Menke,~T.
  \latin{et~al.}  Noisy Intermediate-Scale Quantum Algorithms. \emph{Rev. Mod.
  Phys.} \textbf{2022}, \emph{94}, 015004, DOI:
  \doi{10.1103/RevModPhys.94.015004}\relax
\mciteBstWouldAddEndPuncttrue
\mciteSetBstMidEndSepPunct{\mcitedefaultmidpunct}
{\mcitedefaultendpunct}{\mcitedefaultseppunct}\relax
\EndOfBibitem
\bibitem[Endo \latin{et~al.}(2021)Endo, Cai, Benjamin, and
  Yuan]{Endo.2021.10.7566/JPSJ.90.032001}
Endo,~S.; Cai,~Z.; Benjamin,~S.~C.; Yuan,~X. Hybrid Quantum-Classical
  Algorithms and Quantum Error Mitigation. \emph{J. Phys. Soc. Jpn.}
  \textbf{2021}, \emph{90}, 032001, DOI: \doi{10.7566/JPSJ.90.032001}\relax
\mciteBstWouldAddEndPuncttrue
\mciteSetBstMidEndSepPunct{\mcitedefaultmidpunct}
{\mcitedefaultendpunct}{\mcitedefaultseppunct}\relax
\EndOfBibitem
\bibitem[Callison and Chancellor(2022)Callison, and
  Chancellor]{Callison.2022.10.1103/PhysRevA.106.010101}
Callison,~A.; Chancellor,~N. Hybrid Quantum Classical Algorithms in the Noisy
  Intermediate-Scale Quantum Era and Beyond. \emph{Phys. Rev. A} \textbf{2022},
  \emph{106}, 010101, DOI: \doi{10.1103/PhysRevA.106.010101}\relax
\mciteBstWouldAddEndPuncttrue
\mciteSetBstMidEndSepPunct{\mcitedefaultmidpunct}
{\mcitedefaultendpunct}{\mcitedefaultseppunct}\relax
\EndOfBibitem
\bibitem[Peruzzo \latin{et~al.}(2014)Peruzzo, McClean, Shadbolt, Yung, Zhou,
  Love, Aspuru-Guzik, and O'Brien]{Peruzzo.2014.10.1038/ncomms5213}
Peruzzo,~A.; McClean,~J.; Shadbolt,~P.; Yung,~M.-H.; Zhou,~X.-Q.; Love,~P.~J.;
  Aspuru-Guzik,~A.; O'Brien,~J.~L. A Variational Eigenvalue Solver on a
  Photonic Quantum Processor. \emph{Nat. Commun.} \textbf{2014}, \emph{5},
  4213, DOI: \doi{10.1038/ncomms5213}\relax
\mciteBstWouldAddEndPuncttrue
\mciteSetBstMidEndSepPunct{\mcitedefaultmidpunct}
{\mcitedefaultendpunct}{\mcitedefaultseppunct}\relax
\EndOfBibitem
\bibitem[McClean \latin{et~al.}(2016)McClean, Romero, Babbush, and
  Aspuru-Guzik]{McClean.2016.10.1088/1367-2630/18/2/023023}
McClean,~J.~R.; Romero,~J.; Babbush,~R.; Aspuru-Guzik,~A. The Theory of
  Variational Quantum-Classical Algorithms. \emph{New J. Phys.} \textbf{2016},
  \emph{18}, 023023, DOI: \doi{10.1088/1367-2630/18/2/023023}\relax
\mciteBstWouldAddEndPuncttrue
\mciteSetBstMidEndSepPunct{\mcitedefaultmidpunct}
{\mcitedefaultendpunct}{\mcitedefaultseppunct}\relax
\EndOfBibitem
\bibitem[Cerezo \latin{et~al.}(2021)Cerezo, Arrasmith, Babbush, Benjamin, Endo,
  Fujii, McClean, Mitarai, Yuan, Cincio, and
  Coles]{Cerezo.2021.10.1038/s42254-021-00348-9}
Cerezo,~M.; Arrasmith,~A.; Babbush,~R.; Benjamin,~S.~C.; Endo,~S.; Fujii,~K.;
  McClean,~J.~R.; Mitarai,~K.; Yuan,~X.; Cincio,~L. \latin{et~al.}  Variational
  Quantum Algorithms. \emph{Nat. Rev. Phys.} \textbf{2021}, \emph{3}, 625--644,
  DOI: \doi{10.1038/s42254-021-00348-9}\relax
\mciteBstWouldAddEndPuncttrue
\mciteSetBstMidEndSepPunct{\mcitedefaultmidpunct}
{\mcitedefaultendpunct}{\mcitedefaultseppunct}\relax
\EndOfBibitem
\bibitem[Tilly \latin{et~al.}(2022)Tilly, Chen, Cao, Picozzi, Setia, Li, Grant,
  Wossnig, Rungger, Booth, and
  Tennyson]{Tilly.2022.10.1016/j.physrep.2022.08.003}
Tilly,~J.; Chen,~H.; Cao,~S.; Picozzi,~D.; Setia,~K.; Li,~Y.; Grant,~E.;
  Wossnig,~L.; Rungger,~I.; Booth,~G.~H. \latin{et~al.}  The Variational
  Quantum Eigensolver: A Review of Methods and Best Practices. \emph{Phys.
  Rep.} \textbf{2022}, \emph{986}, 1--128, DOI:
  \doi{10.1016/j.physrep.2022.08.003}\relax
\mciteBstWouldAddEndPuncttrue
\mciteSetBstMidEndSepPunct{\mcitedefaultmidpunct}
{\mcitedefaultendpunct}{\mcitedefaultseppunct}\relax
\EndOfBibitem
\bibitem[Fedorov \latin{et~al.}(2022)Fedorov, Peng, Govind, and
  Alexeev]{Fedorov.2022.10.1186/s41313-021-00032-6}
Fedorov,~D.~A.; Peng,~B.; Govind,~N.; Alexeev,~Y. {VQE} Method: A Short Survey
  and Recent Developments. \emph{Mater. Theory} \textbf{2022}, \emph{6}, 2,
  DOI: \doi{10.1186/s41313-021-00032-6}\relax
\mciteBstWouldAddEndPuncttrue
\mciteSetBstMidEndSepPunct{\mcitedefaultmidpunct}
{\mcitedefaultendpunct}{\mcitedefaultseppunct}\relax
\EndOfBibitem
\bibitem[Smart and Mazziotti(2021)Smart, and
  Mazziotti]{Smart.2021.10.1103/PhysRevLett.126.070504}
Smart,~S.~E.; Mazziotti,~D.~A. Quantum Solver of Contracted Eigenvalue
  Equations for Scalable Molecular Simulations on Quantum Computing Devices.
  \emph{Phys. Rev. Lett.} \textbf{2021}, \emph{126}, 070504, DOI:
  \doi{10.1103/PhysRevLett.126.070504}\relax
\mciteBstWouldAddEndPuncttrue
\mciteSetBstMidEndSepPunct{\mcitedefaultmidpunct}
{\mcitedefaultendpunct}{\mcitedefaultseppunct}\relax
\EndOfBibitem
\bibitem[Stair and Evangelista(2021)Stair, and
  Evangelista]{Stair.2021.10.1103/PRXQuantum.2.030301}
Stair,~N.~H.; Evangelista,~F.~A. Simulating Many-Body Systems with a Projective
  Quantum Eigensolver. \emph{PRX Quantum} \textbf{2021}, \emph{2}, 030301, DOI:
  \doi{10.1103/PRXQuantum.2.030301}\relax
\mciteBstWouldAddEndPuncttrue
\mciteSetBstMidEndSepPunct{\mcitedefaultmidpunct}
{\mcitedefaultendpunct}{\mcitedefaultseppunct}\relax
\EndOfBibitem
\bibitem[Motta \latin{et~al.}(2020)Motta, Sun, Tan, O'Rourke, Ye, Minnich,
  {a}o, and Chan]{Motta.2020.10.1038/s41567-019-0704-4}
Motta,~M.; Sun,~C.; Tan,~A. T.~K.; O'Rourke,~M.~J.; Ye,~E.; Minnich,~A.~J.;
  {a}o,~F. G. S. L.~B.; Chan,~G. K.-L. Determining Eigenstates and Thermal
  States on a Quantum Computer Using Quantum Imaginary Time Evolution.
  \emph{Nat. Phys.} \textbf{2020}, \emph{16}, 205--210, DOI:
  \doi{10.1038/s41567-019-0704-4}\relax
\mciteBstWouldAddEndPuncttrue
\mciteSetBstMidEndSepPunct{\mcitedefaultmidpunct}
{\mcitedefaultendpunct}{\mcitedefaultseppunct}\relax
\EndOfBibitem
\bibitem[Sun \latin{et~al.}(2021)Sun, Motta, Tazhigulov, Tan, Chan, and
  Minnich]{Sun.2021.10.1103/PRXQuantum.2.010317}
Sun,~S.-N.; Motta,~M.; Tazhigulov,~R.~N.; Tan,~A. T.~K.; Chan,~G. K.-L.;
  Minnich,~A.~J. Quantum Computation of Finite-Temperature Static and Dynamical
  Properties of Spin Systems Using Quantum Imaginary Time Evolution. \emph{PRX
  Quantum} \textbf{2021}, \emph{2}, 010317, DOI:
  \doi{10.1103/PRXQuantum.2.010317}\relax
\mciteBstWouldAddEndPuncttrue
\mciteSetBstMidEndSepPunct{\mcitedefaultmidpunct}
{\mcitedefaultendpunct}{\mcitedefaultseppunct}\relax
\EndOfBibitem
\bibitem[McClean \latin{et~al.}(2017)McClean, Kimchi-Schwartz, Carter, and {de
  Jong}]{McClean.2017.10.1103/PhysRevA.95.042308}
McClean,~J.~R.; Kimchi-Schwartz,~M.~E.; Carter,~J.; {de Jong},~W.~A. Hybrid
  Quantum-Classical Hierarchy for Mitigation of Decoherence and Determination
  of Excited States. \emph{Phys. Rev. A} \textbf{2017}, \emph{95}, 042308, DOI:
  \doi{10.1103/PhysRevA.95.042308}\relax
\mciteBstWouldAddEndPuncttrue
\mciteSetBstMidEndSepPunct{\mcitedefaultmidpunct}
{\mcitedefaultendpunct}{\mcitedefaultseppunct}\relax
\EndOfBibitem
\bibitem[Parrish and McMahon()Parrish, and McMahon]{Parrish.2019.1909.08925v1}
Parrish,~R.~M.; McMahon,~P.~L. Quantum Filter Diagonalization: Quantum
  Eigendecomposition without Full Quantum Phase Estimation. 2019,
  arXiv:1909.08925v1. arXiv.org e-Print archive.
  \url{https://arxiv.org/abs/1909.08925v1}\relax
\mciteBstWouldAddEndPuncttrue
\mciteSetBstMidEndSepPunct{\mcitedefaultmidpunct}
{\mcitedefaultendpunct}{\mcitedefaultseppunct}\relax
\EndOfBibitem
\bibitem[Stair \latin{et~al.}(2020)Stair, Huang, and
  Evangelista]{Stair.2020.10.1021/acs.jctc.9b01125}
Stair,~N.~H.; Huang,~R.; Evangelista,~F.~A. A Multireference Quantum Krylov
  Algorithm for Strongly Correlated Electrons. \emph{J. Chem. Theory Comput.}
  \textbf{2020}, \emph{16}, 2236--2245, DOI:
  \doi{10.1021/acs.jctc.9b01125}\relax
\mciteBstWouldAddEndPuncttrue
\mciteSetBstMidEndSepPunct{\mcitedefaultmidpunct}
{\mcitedefaultendpunct}{\mcitedefaultseppunct}\relax
\EndOfBibitem
\bibitem[Huggins \latin{et~al.}(2020)Huggins, Lee, Baek, O'Gorman, and
  Whaley]{Huggins.2020.10.1088/1367-2630/ab867b}
Huggins,~W.~J.; Lee,~J.; Baek,~U.; O'Gorman,~B.; Whaley,~K.~B. A Non-Orthogonal
  Variational Quantum Eigensolver. \emph{New J. Phys.} \textbf{2020},
  \emph{22}, 073009, DOI: \doi{10.1088/1367-2630/ab867b}\relax
\mciteBstWouldAddEndPuncttrue
\mciteSetBstMidEndSepPunct{\mcitedefaultmidpunct}
{\mcitedefaultendpunct}{\mcitedefaultseppunct}\relax
\EndOfBibitem
\bibitem[Kitaev()]{Kitaev.1995.quant-ph/9511026}
Kitaev,~A.~Y. Quantum Measurements and the Abelian Stabilizer Problem. 1995,
  arXiv:quant-ph/9511026. arXiv.org e-Print archive.
  \url{https://arxiv.org/abs/quant-ph/9511026}\relax
\mciteBstWouldAddEndPuncttrue
\mciteSetBstMidEndSepPunct{\mcitedefaultmidpunct}
{\mcitedefaultendpunct}{\mcitedefaultseppunct}\relax
\EndOfBibitem
\bibitem[Abrams and Lloyd(1997)Abrams, and
  Lloyd]{Abrams.1997.10.1103/PhysRevLett.79.2586}
Abrams,~D.~S.; Lloyd,~S. Simulation of Many-Body {Fermi} Systems on a Universal
  Quantum Computer. \emph{Phys. Rev. Lett.} \textbf{1997}, \emph{79},
  2586--2589, DOI: \doi{10.1103/PhysRevLett.79.2586}\relax
\mciteBstWouldAddEndPuncttrue
\mciteSetBstMidEndSepPunct{\mcitedefaultmidpunct}
{\mcitedefaultendpunct}{\mcitedefaultseppunct}\relax
\EndOfBibitem
\bibitem[Abrams and Lloyd(1999)Abrams, and
  Lloyd]{Abrams.1999.10.1103/PhysRevLett.83.5162}
Abrams,~D.~S.; Lloyd,~S. Quantum Algorithm Providing Exponential Speed Increase
  for Finding Eigenvalues and Eigenvectors. \emph{Phys. Rev. Lett.}
  \textbf{1999}, \emph{83}, 5162--5165, DOI:
  \doi{10.1103/PhysRevLett.83.5162}\relax
\mciteBstWouldAddEndPuncttrue
\mciteSetBstMidEndSepPunct{\mcitedefaultmidpunct}
{\mcitedefaultendpunct}{\mcitedefaultseppunct}\relax
\EndOfBibitem
\bibitem[Grimsley \latin{et~al.}(2019)Grimsley, Economou, Barnes, and
  Mayhall]{Grimsley.2019.10.1038/s41467-019-10988-2}
Grimsley,~H.~R.; Economou,~S.~E.; Barnes,~E.; Mayhall,~N.~J. An Adaptive
  Variational Algorithm for Exact Molecular Simulations on a Quantum Computer.
  \emph{Nat Commun.} \textbf{2019}, \emph{10}, 3007, DOI:
  \doi{10.1038/s41467-019-10988-2}\relax
\mciteBstWouldAddEndPuncttrue
\mciteSetBstMidEndSepPunct{\mcitedefaultmidpunct}
{\mcitedefaultendpunct}{\mcitedefaultseppunct}\relax
\EndOfBibitem
\bibitem[Tang \latin{et~al.}(2021)Tang, Shkolnikov, Barron, Grimsley, Mayhall,
  Barnes, and Economou]{Tang.2021.10.1103/PRXQuantum.2.020310}
Tang,~H.~L.; Shkolnikov,~V.~O.; Barron,~G.~S.; Grimsley,~H.~R.; Mayhall,~N.~J.;
  Barnes,~E.; Economou,~S.~E. {Qubit-ADAPT-VQE}: An Adaptive Algorithm for
  Constructing Hardware-Efficient Ans\"{a}tze on a Quantum Processor. \emph{PRX
  Quantum} \textbf{2021}, \emph{2}, 020310, DOI:
  \doi{10.1103/PRXQuantum.2.020310}\relax
\mciteBstWouldAddEndPuncttrue
\mciteSetBstMidEndSepPunct{\mcitedefaultmidpunct}
{\mcitedefaultendpunct}{\mcitedefaultseppunct}\relax
\EndOfBibitem
\bibitem[Yordanov \latin{et~al.}(2021)Yordanov, Armaos, Barnes, and
  Arvidsson-Shukur]{Yordanov.2021.10.1038/s42005-021-00730-0}
Yordanov,~Y.~S.; Armaos,~V.; Barnes,~C. H.~W.; Arvidsson-Shukur,~D. R.~M.
  Qubit-Excitation-Based Adaptive Variational Quantum Eigensolver.
  \emph{Commun. Phys.} \textbf{2021}, \emph{4}, 228, DOI:
  \doi{10.1038/s42005-021-00730-0}\relax
\mciteBstWouldAddEndPuncttrue
\mciteSetBstMidEndSepPunct{\mcitedefaultmidpunct}
{\mcitedefaultendpunct}{\mcitedefaultseppunct}\relax
\EndOfBibitem
\bibitem[Ryabinkin \latin{et~al.}(2020)Ryabinkin, Lang, Genin, and
  Izmaylov]{Ryabinkin.2020.10.1021/acs.jctc.9b01084}
Ryabinkin,~I.~G.; Lang,~R.~A.; Genin,~S.~N.; Izmaylov,~A.~F. Iterative Qubit
  Coupled Cluster Approach with Efficient Screening of Generators. \emph{J.
  Chem. Theory Comput.} \textbf{2020}, \emph{16}, 1055--1063, DOI:
  \doi{10.1021/acs.jctc.9b01084}\relax
\mciteBstWouldAddEndPuncttrue
\mciteSetBstMidEndSepPunct{\mcitedefaultmidpunct}
{\mcitedefaultendpunct}{\mcitedefaultseppunct}\relax
\EndOfBibitem
\bibitem[Mazziotti \latin{et~al.}(2021)Mazziotti, Smart, and
  Mazziotti]{Mazziotti.2021.10.1088/1367-2630/ac3573}
Mazziotti,~D.~A.; Smart,~S.~E.; Mazziotti,~A.~R. Quantum Simulation of
  Molecules without Fermionic Encoding of the Wave Function. \emph{New J.
  Phys.} \textbf{2021}, \emph{23}, 113037, DOI:
  \doi{10.1088/1367-2630/ac3573}\relax
\mciteBstWouldAddEndPuncttrue
\mciteSetBstMidEndSepPunct{\mcitedefaultmidpunct}
{\mcitedefaultendpunct}{\mcitedefaultseppunct}\relax
\EndOfBibitem
\bibitem[Smart and Mazziotti(2022)Smart, and
  Mazziotti]{Smart.2022.10.1103/PhysRevA.105.062424}
Smart,~S.~E.; Mazziotti,~D.~A. Many-Fermion Simulation from the Contracted
  Quantum Eigensolver without Fermionic Encoding of the Wave Function.
  \emph{Phys. Rev. A} \textbf{2022}, \emph{105}, 062424, DOI:
  \doi{10.1103/PhysRevA.105.062424}\relax
\mciteBstWouldAddEndPuncttrue
\mciteSetBstMidEndSepPunct{\mcitedefaultmidpunct}
{\mcitedefaultendpunct}{\mcitedefaultseppunct}\relax
\EndOfBibitem
\bibitem[Fedorov \latin{et~al.}(2022)Fedorov, Alexeev, Gray, and
  Otten]{Fedorov.2022.10.22331/q-2022-05-02-703}
Fedorov,~D.~A.; Alexeev,~Y.; Gray,~S.~K.; Otten,~M.~J. Unitary Selective
  Coupled-Cluster Method. \emph{Quantum} \textbf{2022}, \emph{6}, 703, DOI:
  \doi{10.22331/q-2022-05-02-703}\relax
\mciteBstWouldAddEndPuncttrue
\mciteSetBstMidEndSepPunct{\mcitedefaultmidpunct}
{\mcitedefaultendpunct}{\mcitedefaultseppunct}\relax
\EndOfBibitem
\bibitem[Kutzelnigg(1977)]{Kutzelnigg.1977.10.1007/978-1-4757-0887-5_5}
Kutzelnigg,~W. In \emph{Methods of Electronic Structure Theory};
  Schaefer,~H.~F.,~III, Ed.; Springer: Boston, 1977; pp 129--188, DOI:
  \doi{10.1007/978-1-4757-0887-5_5}\relax
\mciteBstWouldAddEndPuncttrue
\mciteSetBstMidEndSepPunct{\mcitedefaultmidpunct}
{\mcitedefaultendpunct}{\mcitedefaultseppunct}\relax
\EndOfBibitem
\bibitem[Kutzelnigg(1982)]{Kutzelnigg.1982.10.1063/1.444231}
Kutzelnigg,~W. Quantum Chemistry in {Fock} Space. {I.} {The} Universal Wave and
  Energy Operators. \emph{J. Chem. Phys.} \textbf{1982}, \emph{77}, 3081--3097,
  DOI: \doi{10.1063/1.444231}\relax
\mciteBstWouldAddEndPuncttrue
\mciteSetBstMidEndSepPunct{\mcitedefaultmidpunct}
{\mcitedefaultendpunct}{\mcitedefaultseppunct}\relax
\EndOfBibitem
\bibitem[Kutzelnigg and Koch(1983)Kutzelnigg, and
  Koch]{Kutzelnigg.1983.10.1063/1.446313}
Kutzelnigg,~W.; Koch,~S. Quantum Chemistry in {Fock} Space. {II.} {Effective}
  {Hamiltonians} in {Fock} Space. \emph{J. Chem. Phys.} \textbf{1983},
  \emph{79}, 4315--4335, DOI: \doi{10.1063/1.446313}\relax
\mciteBstWouldAddEndPuncttrue
\mciteSetBstMidEndSepPunct{\mcitedefaultmidpunct}
{\mcitedefaultendpunct}{\mcitedefaultseppunct}\relax
\EndOfBibitem
\bibitem[Kutzelnigg(1984)]{Kutzelnigg.1984.10.1063/1.446736}
Kutzelnigg,~W. Quantum Chemistry in {Fock} Space. {III.} {Particle}-Hole
  Formalism. \emph{J. Chem. Phys.} \textbf{1984}, \emph{80}, 822--830, DOI:
  \doi{10.1063/1.446736}\relax
\mciteBstWouldAddEndPuncttrue
\mciteSetBstMidEndSepPunct{\mcitedefaultmidpunct}
{\mcitedefaultendpunct}{\mcitedefaultseppunct}\relax
\EndOfBibitem
\bibitem[Bartlett \latin{et~al.}(1989)Bartlett, Kucharski, and
  Noga]{Bartlett.1989.10.1016/S0009-2614(89)87372-5}
Bartlett,~R.~J.; Kucharski,~S.~A.; Noga,~J. Alternative Coupled-Cluster
  Ans\"{a}tze {II.} {The} Unitary Coupled-Cluster Method. \emph{Chem. Phys.
  Lett.} \textbf{1989}, \emph{155}, 133--140, DOI:
  \doi{10.1016/S0009-2614(89)87372-5}\relax
\mciteBstWouldAddEndPuncttrue
\mciteSetBstMidEndSepPunct{\mcitedefaultmidpunct}
{\mcitedefaultendpunct}{\mcitedefaultseppunct}\relax
\EndOfBibitem
\bibitem[Szalay \latin{et~al.}(1995)Szalay, Nooijen, and
  Bartlett]{Szalay.1995.10.1063/1.469641}
Szalay,~P.~G.; Nooijen,~M.; Bartlett,~R.~J. Alternative Ans\"{a}tze in Single
  Reference Coupled-Cluster Theory. {III.} {A} Critical Analysis of Different
  Methods. \emph{J. Chem. Phys.} \textbf{1995}, \emph{103}, 281--298, DOI:
  \doi{10.1063/1.469641}\relax
\mciteBstWouldAddEndPuncttrue
\mciteSetBstMidEndSepPunct{\mcitedefaultmidpunct}
{\mcitedefaultendpunct}{\mcitedefaultseppunct}\relax
\EndOfBibitem
\bibitem[Taube and Bartlett(2006)Taube, and
  Bartlett]{Taube.2006.10.1002/qua.21198}
Taube,~A.~G.; Bartlett,~R.~J. New Perspectives on Unitary Coupled-Cluster
  Theory. \emph{Int. J. Quantum Chem.} \textbf{2006}, \emph{106}, 3393--3401,
  DOI: \doi{10.1002/qua.21198}\relax
\mciteBstWouldAddEndPuncttrue
\mciteSetBstMidEndSepPunct{\mcitedefaultmidpunct}
{\mcitedefaultendpunct}{\mcitedefaultseppunct}\relax
\EndOfBibitem
\bibitem[Cooper and Knowles(2010)Cooper, and
  Knowles]{Cooper.2010.10.1063/1.3520564}
Cooper,~B.; Knowles,~P.~J. Benchmark Studies of Variational, Unitary and
  Extended Coupled Cluster Methods. \emph{J. Chem. Phys.} \textbf{2010},
  \emph{133}, 234102, DOI: \doi{10.1063/1.3520564}\relax
\mciteBstWouldAddEndPuncttrue
\mciteSetBstMidEndSepPunct{\mcitedefaultmidpunct}
{\mcitedefaultendpunct}{\mcitedefaultseppunct}\relax
\EndOfBibitem
\bibitem[Evangelista(2011)]{Evangelista.2011.10.1063/1.3598471}
Evangelista,~F.~A. Alternative Single-Reference Coupled Cluster Approaches for
  Multireference Problems: The Simpler, the Better. \emph{J. Chem. Phys.}
  \textbf{2011}, \emph{134}, 224102, DOI: \doi{10.1063/1.3598471}\relax
\mciteBstWouldAddEndPuncttrue
\mciteSetBstMidEndSepPunct{\mcitedefaultmidpunct}
{\mcitedefaultendpunct}{\mcitedefaultseppunct}\relax
\EndOfBibitem
\bibitem[Harsha \latin{et~al.}(2018)Harsha, Shiozaki, and
  Scuseria]{Harsha.2018.10.1063/1.5011033}
Harsha,~G.; Shiozaki,~T.; Scuseria,~G.~E. On the Difference Between Variational
  and Unitary Coupled Clutser Theories. \emph{J. Chem. Phys.} \textbf{2018},
  \emph{148}, 044107, DOI: \doi{10.1063/1.5011033}\relax
\mciteBstWouldAddEndPuncttrue
\mciteSetBstMidEndSepPunct{\mcitedefaultmidpunct}
{\mcitedefaultendpunct}{\mcitedefaultseppunct}\relax
\EndOfBibitem
\bibitem[Filip and Thom(2020)Filip, and Thom]{Filip.2020.10.1063/5.0026141}
Filip,~M.-A.; Thom,~A. J.~W. A Stochastic Approach to Unitary Coupled Cluster.
  \emph{J. Chem. Phys.} \textbf{2020}, \emph{153}, 214106, DOI:
  \doi{10.1063/5.0026141}\relax
\mciteBstWouldAddEndPuncttrue
\mciteSetBstMidEndSepPunct{\mcitedefaultmidpunct}
{\mcitedefaultendpunct}{\mcitedefaultseppunct}\relax
\EndOfBibitem
\bibitem[Freericks(2022)]{Freericks.2022.10.3390/sym14030494}
Freericks,~J.~K. Operator Relationship between Conventional Coupled Cluster and
  Unitary Coupled Cluster. \emph{Symmetry} \textbf{2022}, \emph{14}, 494, DOI:
  \doi{10.3390/sym14030494}\relax
\mciteBstWouldAddEndPuncttrue
\mciteSetBstMidEndSepPunct{\mcitedefaultmidpunct}
{\mcitedefaultendpunct}{\mcitedefaultseppunct}\relax
\EndOfBibitem
\bibitem[Anand \latin{et~al.}(2022)Anand, Schleich, Alperin-Lea, Jensen, Sim,
  Diaz-Tinoco, Kottmann, Degroote, Izmaylov, and
  Aspuru-Guzik]{Anand.2022.10.1039/d1cs00932j}
Anand,~A.; Schleich,~P.; Alperin-Lea,~S.; Jensen,~P. W.~K.; Sim,~S.;
  Diaz-Tinoco,~M.; Kottmann,~J.~S.; Degroote,~M.; Izmaylov,~A.~F.;
  Aspuru-Guzik,~A. A Quantum Computing View on Unitary Coupled Cluster Theory.
  \emph{Chem. Soc. Rev.} \textbf{2022}, \emph{51}, 1659--1684, DOI:
  \doi{10.1039/d1cs00932j}\relax
\mciteBstWouldAddEndPuncttrue
\mciteSetBstMidEndSepPunct{\mcitedefaultmidpunct}
{\mcitedefaultendpunct}{\mcitedefaultseppunct}\relax
\EndOfBibitem
\bibitem[Coester(1958)]{Coester.1958.10.1016/0029-5582(58)90280-3}
Coester,~F. Bound States of a Many-Particle System. \emph{Nucl. Phys.}
  \textbf{1958}, \emph{7}, 421--424, DOI:
  \doi{10.1016/0029-5582(58)90280-3}\relax
\mciteBstWouldAddEndPuncttrue
\mciteSetBstMidEndSepPunct{\mcitedefaultmidpunct}
{\mcitedefaultendpunct}{\mcitedefaultseppunct}\relax
\EndOfBibitem
\bibitem[Coester and K{\" u}mmel(1960)Coester, and K{\"
  u}mmel]{Coester.1960.10.1016/0029-5582(60)90140-1}
Coester,~F.; K{\" u}mmel,~H. Short-Range Correlations in Nuclear Wave
  Functions. \emph{Nucl. Phys.} \textbf{1960}, \emph{17}, 477--485, DOI:
  \doi{10.1016/0029-5582(60)90140-1}\relax
\mciteBstWouldAddEndPuncttrue
\mciteSetBstMidEndSepPunct{\mcitedefaultmidpunct}
{\mcitedefaultendpunct}{\mcitedefaultseppunct}\relax
\EndOfBibitem
\bibitem[{\v C}{\'\i}{\v z}ek(1966)]{Cizek.1966.10.1063/1.1727484}
{\v C}{\'\i}{\v z}ek,~J. On the Correlation Problem in Atomic and Molecular
  Systems. {Calculation} of Wavefunction Components in {Ursell}-Type Expansion
  Using Quantum-Field Theoretical Methods. \emph{J. Chem. Phys.} \textbf{1966},
  \emph{45}, 4256--4266, DOI: \doi{10.1063/1.1727484}\relax
\mciteBstWouldAddEndPuncttrue
\mciteSetBstMidEndSepPunct{\mcitedefaultmidpunct}
{\mcitedefaultendpunct}{\mcitedefaultseppunct}\relax
\EndOfBibitem
\bibitem[{\v C}{\'\i}{\v z}ek(1969)]{Cizek.1969.10.1002/9780470143599.ch2}
{\v C}{\'\i}{\v z}ek,~J. On the Use of the Cluster Expansion and the Technique
  of Diagrams in Calculations of Correlation Effects in Atoms and Molecules.
  \emph{Adv. Chem. Phys.} \textbf{1969}, \emph{14}, 35--89, DOI:
  \doi{10.1002/9780470143599.ch2}\relax
\mciteBstWouldAddEndPuncttrue
\mciteSetBstMidEndSepPunct{\mcitedefaultmidpunct}
{\mcitedefaultendpunct}{\mcitedefaultseppunct}\relax
\EndOfBibitem
\bibitem[{\v C}{\'\i}{\v z}ek and Paldus(1971){\v C}{\'\i}{\v z}ek, and
  Paldus]{Cizek.1971.10.1002/qua.560050402}
{\v C}{\'\i}{\v z}ek,~J.; Paldus,~J. Correlation Problems in Atomic and
  Molecular Systems {III.} {Rederivation} of the Coupled-Pair Many-Electron
  Theory Using the Traditional Quantum Chemical Methods. \emph{Int. J. Quantum
  Chem.} \textbf{1971}, \emph{5}, 359--379, DOI:
  \doi{10.1002/qua.560050402}\relax
\mciteBstWouldAddEndPuncttrue
\mciteSetBstMidEndSepPunct{\mcitedefaultmidpunct}
{\mcitedefaultendpunct}{\mcitedefaultseppunct}\relax
\EndOfBibitem
\bibitem[Paldus \latin{et~al.}(1972)Paldus, {\v C}{\'\i}{\v z}ek, and
  Shavitt]{Paldus.1972.10.1103/PhysRevA.5.50}
Paldus,~J.; {\v C}{\'\i}{\v z}ek,~J.; Shavitt,~I. Correlation Problems in
  Atomic and Molecular Systems. {IV.} {Extended} Coupled-Pair Many-Electron
  Theory and Its Application to the $\text{BH}_3$ Molecule. \emph{Phys. Rev. A}
  \textbf{1972}, \emph{5}, 50--67, DOI: \doi{10.1103/PhysRevA.5.50}\relax
\mciteBstWouldAddEndPuncttrue
\mciteSetBstMidEndSepPunct{\mcitedefaultmidpunct}
{\mcitedefaultendpunct}{\mcitedefaultseppunct}\relax
\EndOfBibitem
\bibitem[Ryabinkin \latin{et~al.}(2018)Ryabinkin, Yen, Genin, and
  Izmaylov]{Ryabinkin.2018.10.1021/acs.jctc.8b00932}
Ryabinkin,~I.~G.; Yen,~T.-C.; Genin,~S.~N.; Izmaylov,~A.~F. Qubit Coupled
  Cluster Method: A Systematic Approach to Quantum Chemistry on a Quantum
  Computer. \emph{J. Chem. Theory Comput.} \textbf{2018}, \emph{14},
  6317--6326, DOI: \doi{10.1021/acs.jctc.8b00932}\relax
\mciteBstWouldAddEndPuncttrue
\mciteSetBstMidEndSepPunct{\mcitedefaultmidpunct}
{\mcitedefaultendpunct}{\mcitedefaultseppunct}\relax
\EndOfBibitem
\bibitem[Yordanov \latin{et~al.}(2020)Yordanov, Arvidsson-Shukur, and
  Barnes]{Yordanov.2020.10.1103/PhysRevA.102.062612}
Yordanov,~Y.~S.; Arvidsson-Shukur,~D. R.~M.; Barnes,~C. H.~W. Efficient Quantum
  Circuits for Quantum Computational Chemistry. \emph{Phys. Rev. A}
  \textbf{2020}, \emph{102}, 062612, DOI:
  \doi{10.1103/PhysRevA.102.062612}\relax
\mciteBstWouldAddEndPuncttrue
\mciteSetBstMidEndSepPunct{\mcitedefaultmidpunct}
{\mcitedefaultendpunct}{\mcitedefaultseppunct}\relax
\EndOfBibitem
\bibitem[Xia and Kais(2021)Xia, and Kais]{Xia.2021.10.1088/2058-9565/abbc74}
Xia,~R.; Kais,~S. Qubit Coupled Cluster Singles and Doubles Variational Quantum
  Eigensolver Ansatz for Electronic Structure Calculations. \emph{Quantum Sci.
  Technol.} \textbf{2021}, \emph{6}, 015001, DOI:
  \doi{10.1088/2058-9565/abbc74}\relax
\mciteBstWouldAddEndPuncttrue
\mciteSetBstMidEndSepPunct{\mcitedefaultmidpunct}
{\mcitedefaultendpunct}{\mcitedefaultseppunct}\relax
\EndOfBibitem
\bibitem[Magoulas and Evangelista(2023)Magoulas, and
  Evangelista]{Magoulas.2023.10.1021/acs.jctc.2c01016}
Magoulas,~I.; Evangelista,~F.~A. {CNOT}-Efficient Quantum Circuits for
  Arbitrary Rank Many-Body Fermionic and Qubit Excitations. \emph{J. Chem.
  Theory Comput.} \textbf{2023}, \emph{19}, 822--836, DOI:
  \doi{10.1021/acs.jctc.2c01016}\relax
\mciteBstWouldAddEndPuncttrue
\mciteSetBstMidEndSepPunct{\mcitedefaultmidpunct}
{\mcitedefaultendpunct}{\mcitedefaultseppunct}\relax
\EndOfBibitem
\bibitem[Magoulas and Evangelista()Magoulas, and
  Evangelista]{Magoulas.2023.2304.12870}
Magoulas,~I.; Evangelista,~F.~A. Linear-Scaling Quantumm Circuits for
  Computational Chemistry. 2023, arXiv:2304.12870. arXiv.org e-Print archive.
  \url{https://arxiv.org/abs/2304.12870}\relax
\mciteBstWouldAddEndPuncttrue
\mciteSetBstMidEndSepPunct{\mcitedefaultmidpunct}
{\mcitedefaultendpunct}{\mcitedefaultseppunct}\relax
\EndOfBibitem
\bibitem[Ryabinkin \latin{et~al.}(2021)Ryabinkin, Izmaylov, and
  Genin]{Ryabinkin.2021.10.1088/2058-9565/abda8e}
Ryabinkin,~I.~G.; Izmaylov,~A.~F.; Genin,~S.~N. \textit{A Posteriori}
  Corrections to the Iterative Qubit Coupled Cluster Method to Minimize the Use
  of Quantum Resources in Large-Scale Calculations. \emph{Quantum Sci.
  Technol.} \textbf{2021}, \emph{6}, 024012, DOI:
  \doi{10.1088/2058-9565/abda8e}\relax
\mciteBstWouldAddEndPuncttrue
\mciteSetBstMidEndSepPunct{\mcitedefaultmidpunct}
{\mcitedefaultendpunct}{\mcitedefaultseppunct}\relax
\EndOfBibitem
\bibitem[Piecuch and Kowalski(2000)Piecuch, and
  Kowalski]{Piecuch.2000.10.1142/9789812792501_0001}
Piecuch,~P.; Kowalski,~K. In \emph{Computational Chemistry: Reviews of Current
  Trends}; Leszczy{\' n}ski,~J., Ed.; World Scientific: Singapore, 2000;
  Vol.~5; pp 1--104, DOI: \doi{10.1142/9789812792501_0001}\relax
\mciteBstWouldAddEndPuncttrue
\mciteSetBstMidEndSepPunct{\mcitedefaultmidpunct}
{\mcitedefaultendpunct}{\mcitedefaultseppunct}\relax
\EndOfBibitem
\bibitem[Kowalski and Piecuch(2000)Kowalski, and
  Piecuch]{Kowalski.2000.10.1063/1.481769}
Kowalski,~K.; Piecuch,~P. The Method of Moments of Coupled-Cluster Equations
  and the Renormalized {CCSD[T]}, {CCSD(T)}, {CCSD(TQ)}, and {CCSDT(Q)}
  Approaches. \emph{J. Chem. Phys.} \textbf{2000}, \emph{113}, 18--35, DOI:
  \doi{10.1063/1.481769}\relax
\mciteBstWouldAddEndPuncttrue
\mciteSetBstMidEndSepPunct{\mcitedefaultmidpunct}
{\mcitedefaultendpunct}{\mcitedefaultseppunct}\relax
\EndOfBibitem
\bibitem[Kowalski and Piecuch(2000)Kowalski, and
  Piecuch]{Kowalski.2000.10.1063/1.1290609}
Kowalski,~K.; Piecuch,~P. Renormalized {CCSD(T)} and {CCSD(TQ)} Approaches:
  Dissociation of the $\text{N}_2$ Triple Bond. \emph{J. Chem. Phys.}
  \textbf{2000}, \emph{113}, 5644--5652, DOI: \doi{10.1063/1.1290609}\relax
\mciteBstWouldAddEndPuncttrue
\mciteSetBstMidEndSepPunct{\mcitedefaultmidpunct}
{\mcitedefaultendpunct}{\mcitedefaultseppunct}\relax
\EndOfBibitem
\bibitem[Piecuch \latin{et~al.}(2001)Piecuch, Kucharski, and
  Kowalski]{Piecuch.2001.10.1016/S0009-2614(01)00759-X}
Piecuch,~P.; Kucharski,~S.~A.; Kowalski,~K. Can Ordinary Single-Reference
  Coupled-Cluster Methods Describe the Potential Energy Curve of $\text{N}_2$?
  The Renormalized {CCSDT(Q)} Study. \emph{Chem. Phys. Lett.} \textbf{2001},
  \emph{344}, 176--184, DOI: \doi{10.1016/S0009-2614(01)00759-X}\relax
\mciteBstWouldAddEndPuncttrue
\mciteSetBstMidEndSepPunct{\mcitedefaultmidpunct}
{\mcitedefaultendpunct}{\mcitedefaultseppunct}\relax
\EndOfBibitem
\bibitem[Fan \latin{et~al.}(2005)Fan, Kowalski, and
  Piecuch]{Fan.2005.10.1080/00268970500131595}
Fan,~P.-D.; Kowalski,~K.; Piecuch,~P. Non-Iterative Corrections to Extended
  Coupled-Cluster Energies Employing the Generalized Method of Moments of
  Coupled-Cluster Equations. \emph{Mol. Phys.} \textbf{2005}, \emph{103},
  2191--2213, DOI: \doi{10.1080/00268970500131595}\relax
\mciteBstWouldAddEndPuncttrue
\mciteSetBstMidEndSepPunct{\mcitedefaultmidpunct}
{\mcitedefaultendpunct}{\mcitedefaultseppunct}\relax
\EndOfBibitem
\bibitem[Lodriguito \latin{et~al.}(2006)Lodriguito, Kowalski, W{\l}och, and
  Piecuch]{Lodriguito.2006.10.1016/j.theochem.2006.03.014}
Lodriguito,~M.~D.; Kowalski,~K.; W{\l}och,~M.; Piecuch,~P. Non-Iterative
  Coupled-Cluster Methods Employing Multi-Reference Perturbation Theory Wave
  Functions. \emph{J. Mol. Struct.: THEOCHEM} \textbf{2006}, \emph{771},
  89--104, DOI: \doi{10.1016/j.theochem.2006.03.014}\relax
\mciteBstWouldAddEndPuncttrue
\mciteSetBstMidEndSepPunct{\mcitedefaultmidpunct}
{\mcitedefaultendpunct}{\mcitedefaultseppunct}\relax
\EndOfBibitem
\bibitem[Piecuch and W{\l}och(2005)Piecuch, and
  W{\l}och]{Piecuch.2005.10.1063/1.2137318}
Piecuch,~P.; W{\l}och,~M. Renormalized Coupled-Cluster Methods Exploiting Left
  Eigenstates of the Similarity-Transformed {Hamiltonian}. \emph{J. Chem.
  Phys.} \textbf{2005}, \emph{123}, 224105, DOI: \doi{10.1063/1.2137318}\relax
\mciteBstWouldAddEndPuncttrue
\mciteSetBstMidEndSepPunct{\mcitedefaultmidpunct}
{\mcitedefaultendpunct}{\mcitedefaultseppunct}\relax
\EndOfBibitem
\bibitem[Piecuch \latin{et~al.}(2006)Piecuch, W{\l}och, Gour, and
  Kinal]{Piecuch.2006.10.1016/j.cplett.2005.10.116}
Piecuch,~P.; W{\l}och,~M.; Gour,~J.~R.; Kinal,~A. Single-Reference,
  Size-Extensive, Non-Iterative Coupled-Cluster Approaches to Bond Breaking and
  Biradicals. \emph{Chem. Phys. Lett.} \textbf{2006}, \emph{418}, 467--474,
  DOI: \doi{10.1016/j.cplett.2005.10.116}\relax
\mciteBstWouldAddEndPuncttrue
\mciteSetBstMidEndSepPunct{\mcitedefaultmidpunct}
{\mcitedefaultendpunct}{\mcitedefaultseppunct}\relax
\EndOfBibitem
\bibitem[W{\l}och \latin{et~al.}(2006)W{\l}och, Lodriguito, Piecuch, and
  Gour]{Wloch.2006.10.1080/00268970600659586}
W{\l}och,~M.; Lodriguito,~M.~D.; Piecuch,~P.; Gour,~J.~R. Two New Classes of
  Non-Iterative Coupled-Cluster Methods Derived from the Method of Moments of
  Coupled-Cluster Equations. \emph{Mol. Phys.} \textbf{2006}, \emph{104},
  2149--2172, DOI: \doi{10.1080/00268970600659586}, \textit{ibid.}\
  \textbf{124}, 2291 (2006) [Erratum]\relax
\mciteBstWouldAddEndPuncttrue
\mciteSetBstMidEndSepPunct{\mcitedefaultmidpunct}
{\mcitedefaultendpunct}{\mcitedefaultseppunct}\relax
\EndOfBibitem
\bibitem[W{\l}och \latin{et~al.}(2007)W{\l}och, Gour, and
  Piecuch]{Wloch.2007.10.1021/jp072535l}
W{\l}och,~M.; Gour,~J.~R.; Piecuch,~P. Extension of the Renormalized
  Coupled-Cluster Methods Exploiting Left Eigenstates of the
  Similarity-Transformed {Hamiltonian} to Open-Shell Systems: A Benchmark
  Study. \emph{J. Phys. Chem. A} \textbf{2007}, \emph{111}, 11359--11382, DOI:
  \doi{10.1021/jp072535l}\relax
\mciteBstWouldAddEndPuncttrue
\mciteSetBstMidEndSepPunct{\mcitedefaultmidpunct}
{\mcitedefaultendpunct}{\mcitedefaultseppunct}\relax
\EndOfBibitem
\bibitem[Piecuch \latin{et~al.}(2009)Piecuch, Gour, and
  W{\l}och]{Piecuch.2009.10.1002/qua.22367}
Piecuch,~P.; Gour,~J.~R.; W{\l}och,~M. Left-Eigenstate Completely Renormalized
  Equation-of-Motion Coupled-Cluster Methods: Review of Key Concepts, Extension
  to Excited States of Open-Shell Systems, and Comparison with
  Electron-Attached and Ionized Approaches. \emph{Int. J. Quantum Chem.}
  \textbf{2009}, \emph{109}, 3268--3304, DOI: \doi{10.1002/qua.22367}\relax
\mciteBstWouldAddEndPuncttrue
\mciteSetBstMidEndSepPunct{\mcitedefaultmidpunct}
{\mcitedefaultendpunct}{\mcitedefaultseppunct}\relax
\EndOfBibitem
\bibitem[Piecuch \latin{et~al.}(2002)Piecuch, Kowalski, Pimienta, and
  Mcguire]{Piecuch.2002.10.1080/0144235021000053811}
Piecuch,~P.; Kowalski,~K.; Pimienta,~I. S.~O.; Mcguire,~M.~J. Recent Advances
  in Electronic Structure Theory: Method of Moments of Coupled-Cluster
  Equations and Renormalized Coupled-Cluster Approaches. \emph{Int. Rev. Phys.
  Chem.} \textbf{2002}, \emph{21}, 527--655, DOI:
  \doi{10.1080/0144235021000053811}\relax
\mciteBstWouldAddEndPuncttrue
\mciteSetBstMidEndSepPunct{\mcitedefaultmidpunct}
{\mcitedefaultendpunct}{\mcitedefaultseppunct}\relax
\EndOfBibitem
\bibitem[Piecuch \latin{et~al.}(2004)Piecuch, Kowalski, Pimienta, Fan,
  Lodriguito, McGuire, Kucharski, Ku{\' s}, and
  Musia{\l}]{Piecuch.2004.10.1007/s00214-004-0567-2}
Piecuch,~P.; Kowalski,~K.; Pimienta,~I. S.~O.; Fan,~P.-D.; Lodriguito,~M.;
  McGuire,~M.~J.; Kucharski,~S.~A.; Ku{\' s},~T.; Musia{\l},~M. Method of
  Moments of Coupled-Cluster Equations: A New Formalism for Designing Accurate
  Electronic Structure Methods for Ground and Excited States. \emph{Theor.
  Chem. Acc.} \textbf{2004}, \emph{112}, 349--393, DOI:
  \doi{10.1007/s00214-004-0567-2}\relax
\mciteBstWouldAddEndPuncttrue
\mciteSetBstMidEndSepPunct{\mcitedefaultmidpunct}
{\mcitedefaultendpunct}{\mcitedefaultseppunct}\relax
\EndOfBibitem
\bibitem[Shen and Piecuch(2012)Shen, and
  Piecuch]{Shen.2012.10.1016/j.chemphys.2011.11.033}
Shen,~J.; Piecuch,~P. Biorthogonal Moment Expansions in Coupled-Cluster Theory:
  Review of Key Concepts and Merging the Renormalized and Active-Space
  Coupled-Cluster Methods. \emph{Chem. Phys.} \textbf{2012}, \emph{401},
  180--202, DOI: \doi{10.1016/j.chemphys.2011.11.033}\relax
\mciteBstWouldAddEndPuncttrue
\mciteSetBstMidEndSepPunct{\mcitedefaultmidpunct}
{\mcitedefaultendpunct}{\mcitedefaultseppunct}\relax
\EndOfBibitem
\bibitem[Urban \latin{et~al.}(1985)Urban, Noga, Cole, and
  Bartlett]{Urban.1985.10.1063/1.449067}
Urban,~M.; Noga,~J.; Cole,~S.~J.; Bartlett,~R.~J. Towards a Full {CCSDT} Model
  for Electron Correlation. \emph{J. Chem. Phys.} \textbf{1985}, \emph{83},
  4041--4046, DOI: \doi{10.1063/1.449067}, \textit{ibid.}\ \textbf{85}, 5383
  (1986) [Erratum]\relax
\mciteBstWouldAddEndPuncttrue
\mciteSetBstMidEndSepPunct{\mcitedefaultmidpunct}
{\mcitedefaultendpunct}{\mcitedefaultseppunct}\relax
\EndOfBibitem
\bibitem[Raghavachari(1985)]{Raghavachari.1985.10.1063/1.448718}
Raghavachari,~K. An Augmented Coupled Cluster Method and Its Application to the
  First-Row Homonuclear Diatomics. \emph{J. Chem. Phys.} \textbf{1985},
  \emph{82}, 4607--4610, DOI: \doi{10.1063/1.448718}\relax
\mciteBstWouldAddEndPuncttrue
\mciteSetBstMidEndSepPunct{\mcitedefaultmidpunct}
{\mcitedefaultendpunct}{\mcitedefaultseppunct}\relax
\EndOfBibitem
\bibitem[Raghavachari \latin{et~al.}(1989)Raghavachari, Trucks, Pople, and
  Head-Gordon]{Raghavachari.1989.10.1016/S0009-2614(89)87395-6}
Raghavachari,~K.; Trucks,~G.~W.; Pople,~J.~A.; Head-Gordon,~M. A Fifth-Order
  Perturbation Comparison of Electron Correlation Theories. \emph{Chem. Phys.
  Lett.} \textbf{1989}, \emph{157}, 479--483, DOI:
  \doi{10.1016/S0009-2614(89)87395-6}\relax
\mciteBstWouldAddEndPuncttrue
\mciteSetBstMidEndSepPunct{\mcitedefaultmidpunct}
{\mcitedefaultendpunct}{\mcitedefaultseppunct}\relax
\EndOfBibitem
\bibitem[Kucharski and Bartlett(1989)Kucharski, and
  Bartlett]{Kucharski.1989.10.1016/0009-2614(89)87388-9}
Kucharski,~S.~A.; Bartlett,~R.~J. Coupled-Cluster Methods that Include
  Connected Quadruple Excitations, $T_4$: {CCSDTQ-1} and {Q(CCSDT)}.
  \emph{Chem. Phys. Lett.} \textbf{1989}, \emph{158}, 550--555, DOI:
  \doi{10.1016/0009-2614(89)87388-9}\relax
\mciteBstWouldAddEndPuncttrue
\mciteSetBstMidEndSepPunct{\mcitedefaultmidpunct}
{\mcitedefaultendpunct}{\mcitedefaultseppunct}\relax
\EndOfBibitem
\bibitem[Bartlett \latin{et~al.}(1990)Bartlett, Watts, Kucharski, and
  Noga]{Bartlett.1990.10.1016/0009-2614(90)87031-L}
Bartlett,~R.~J.; Watts,~J.~D.; Kucharski,~S.~A.; Noga,~J. Non-Iterative
  Fifth-Order Triple and Quadruple Excitation Energy Corrections in Correlated
  Methods. \emph{Chem. Phys. Lett.} \textbf{1990}, \emph{165}, 513--522, DOI:
  \doi{10.1016/0009-2614(90)87031-L}\relax
\mciteBstWouldAddEndPuncttrue
\mciteSetBstMidEndSepPunct{\mcitedefaultmidpunct}
{\mcitedefaultendpunct}{\mcitedefaultseppunct}\relax
\EndOfBibitem
\bibitem[Kucharski and Bartlett(1993)Kucharski, and
  Bartlett]{Kucharski.1993.10.1016/0009-2614(93)80186-S}
Kucharski,~S.~A.; Bartlett,~R.~J. Coupled-Cluster Methods Correct trhough Sixth
  Order. \emph{Chem. Phys. Lett.} \textbf{1993}, \emph{206}, 574--583, DOI:
  \doi{10.1016/0009-2614(93)80186-S}\relax
\mciteBstWouldAddEndPuncttrue
\mciteSetBstMidEndSepPunct{\mcitedefaultmidpunct}
{\mcitedefaultendpunct}{\mcitedefaultseppunct}\relax
\EndOfBibitem
\bibitem[Kucharski and Bartlett(1998)Kucharski, and
  Bartlett]{Kucharski.1998.10.1063/1.475961}
Kucharski,~S.~A.; Bartlett,~R.~J. Noniterative Energy Corrections Through
  Fifth-Order to the Coupled Cluster Singles and Doubles Method. \emph{J. Chem.
  Phys.} \textbf{1998}, \emph{108}, 5243--5254, DOI:
  \doi{10.1063/1.475961}\relax
\mciteBstWouldAddEndPuncttrue
\mciteSetBstMidEndSepPunct{\mcitedefaultmidpunct}
{\mcitedefaultendpunct}{\mcitedefaultseppunct}\relax
\EndOfBibitem
\bibitem[Kucharski and Bartlett(1998)Kucharski, and
  Bartlett]{Kucharski.1998.10.1063/1.475962}
Kucharski,~S.~A.; Bartlett,~R.~J. Sixth-Order Energy Corrections with Converged
  Coupled Cluster Singles and Doubles Amplitudes. \emph{J. Chem. Phys.}
  \textbf{1998}, \emph{108}, 5255--5264, DOI: \doi{10.1063/1.475962}\relax
\mciteBstWouldAddEndPuncttrue
\mciteSetBstMidEndSepPunct{\mcitedefaultmidpunct}
{\mcitedefaultendpunct}{\mcitedefaultseppunct}\relax
\EndOfBibitem
\bibitem[Kucharski and Bartlett(1998)Kucharski, and
  Bartlett]{Kucharski.1998.10.1063/1.476376}
Kucharski,~S.~A.; Bartlett,~R.~J. An Efficient Way to Include Connected
  Quadruple Contributions into the Coupled Cluster Method. \emph{J. Chem.
  Phys.} \textbf{1998}, \emph{108}, 9221--9226, DOI:
  \doi{10.1063/1.476376}\relax
\mciteBstWouldAddEndPuncttrue
\mciteSetBstMidEndSepPunct{\mcitedefaultmidpunct}
{\mcitedefaultendpunct}{\mcitedefaultseppunct}\relax
\EndOfBibitem
\bibitem[Musial \latin{et~al.}(2000)Musial, Kucharski, and
  Bartlett]{Musial.2000.10.1016/S0009-2614(00)00290-6}
Musial,~M.; Kucharski,~S.~A.; Bartlett,~R.~J. $T_5$ Operator in Coupled Cluster
  Calculations. \emph{Chem. Phys. Lett.} \textbf{2000}, \emph{320}, 542--548,
  DOI: \doi{10.1016/S0009-2614(00)00290-6}\relax
\mciteBstWouldAddEndPuncttrue
\mciteSetBstMidEndSepPunct{\mcitedefaultmidpunct}
{\mcitedefaultendpunct}{\mcitedefaultseppunct}\relax
\EndOfBibitem
\bibitem[Peng and Kowalski(2022)Peng, and
  Kowalski]{Peng.2022.10.1103/PhysRevResearch.4.043172}
Peng,~B.; Kowalski,~K. Mapping Renormalized Coupled Cluster Methods to Quantum
  Computers through a Compact Unitary Representation of Nonunitary Operators.
  \emph{Phys. Rev. Research} \textbf{2022}, \emph{4}, 043172, DOI:
  \doi{10.1103/PhysRevResearch.4.043172}\relax
\mciteBstWouldAddEndPuncttrue
\mciteSetBstMidEndSepPunct{\mcitedefaultmidpunct}
{\mcitedefaultendpunct}{\mcitedefaultseppunct}\relax
\EndOfBibitem
\bibitem[Schuld \latin{et~al.}(2019)Schuld, Bergholm, Gogolin, Izaac, and
  Killoran]{Schuld.2019.10.1103/PhysRevA.99.032331}
Schuld,~M.; Bergholm,~V.; Gogolin,~C.; Izaac,~J.; Killoran,~N. Evaluating
  Analytic Gradients on Quantum Hardware. \emph{Phys. Rev. A} \textbf{2019},
  \emph{99}, 032331, DOI: \doi{10.1103/PhysRevA.99.032331}\relax
\mciteBstWouldAddEndPuncttrue
\mciteSetBstMidEndSepPunct{\mcitedefaultmidpunct}
{\mcitedefaultendpunct}{\mcitedefaultseppunct}\relax
\EndOfBibitem
\bibitem[Kottmann \latin{et~al.}(2021)Kottmann, Anand, and
  Aspuru-Guzik]{Kottmann.2021.10.1039/d0sc06627c}
Kottmann,~J.~S.; Anand,~A.; Aspuru-Guzik,~A. A Feasible Approach for
  Automatically Differentiable Unitary Coupled-Cluster on Quantum Computers.
  \emph{Chem. Sci.} \textbf{2021}, \emph{12}, 3497--3508, DOI:
  \doi{10.1039/d0sc06627c}\relax
\mciteBstWouldAddEndPuncttrue
\mciteSetBstMidEndSepPunct{\mcitedefaultmidpunct}
{\mcitedefaultendpunct}{\mcitedefaultseppunct}\relax
\EndOfBibitem
\bibitem[Jankowski \latin{et~al.}(1991)Jankowski, Paldus, and
  Piecuch]{Jankowski.1991.10.1007/BF01117411}
Jankowski,~K.; Paldus,~J.; Piecuch,~P. Method of Moments Approach and Coupled
  Cluster Theory. \emph{Theor. Chim. Acta} \textbf{1991}, \emph{80}, 223--243,
  DOI: \doi{10.1007/BF01117411}\relax
\mciteBstWouldAddEndPuncttrue
\mciteSetBstMidEndSepPunct{\mcitedefaultmidpunct}
{\mcitedefaultendpunct}{\mcitedefaultseppunct}\relax
\EndOfBibitem
\bibitem[Shen and Piecuch(2012)Shen, and Piecuch]{Shen.2012.10.1063/1.3700802}
Shen,~J.; Piecuch,~P. Combining Active-Space Coupled-Cluster Methods with
  Moment Energy Corrections via the CC($P$;$Q$) Methodology, with Benchmark
  Calculations for Biradical Transition States. \emph{J. Chem. Phys.}
  \textbf{2012}, \emph{136}, 144104, DOI: \doi{10.1063/1.3700802}\relax
\mciteBstWouldAddEndPuncttrue
\mciteSetBstMidEndSepPunct{\mcitedefaultmidpunct}
{\mcitedefaultendpunct}{\mcitedefaultseppunct}\relax
\EndOfBibitem
\bibitem[Shen and Piecuch(2012)Shen, and Piecuch]{Shen.2012.10.1021/ct300762m}
Shen,~J.; Piecuch,~P. Merging Active-Space and Renormalized Coupled-Cluster
  Methods via the CC($P$;$Q$) Formalism, with Benchmark Calculations for
  Singlet--Triplet Gaps in Biradical Systems. \emph{J. Chem. Theory Comput.}
  \textbf{2012}, \emph{8}, 4968--4988, DOI: \doi{10.1021/ct300762m}\relax
\mciteBstWouldAddEndPuncttrue
\mciteSetBstMidEndSepPunct{\mcitedefaultmidpunct}
{\mcitedefaultendpunct}{\mcitedefaultseppunct}\relax
\EndOfBibitem
\bibitem[Stanton(1997)]{Stanton.1997.10.1016/S0009-2614(97)01144-5}
Stanton,~J.~F. Why {CCSD(T)} Works: A Different Perspective. \emph{Chem. Phys.
  Lett.} \textbf{1997}, \emph{281}, 130--134, DOI:
  \doi{10.1016/S0009-2614(97)01144-5}\relax
\mciteBstWouldAddEndPuncttrue
\mciteSetBstMidEndSepPunct{\mcitedefaultmidpunct}
{\mcitedefaultendpunct}{\mcitedefaultseppunct}\relax
\EndOfBibitem
\bibitem[Huron \latin{et~al.}(1973)Huron, Malrieu, and
  Rancurel]{Huron.1973.10.1063/1.1679199}
Huron,~B.; Malrieu,~J.-P.; Rancurel,~P. Iterative Perturbation Calculations of
  Ground and Excited State Energies from Multiconfigurational Zeroth-Order
  Wavefunctions. \emph{J. Chem. Phys.} \textbf{1973}, \emph{58}, 5745--5759,
  DOI: \doi{10.1063/1.1679199}\relax
\mciteBstWouldAddEndPuncttrue
\mciteSetBstMidEndSepPunct{\mcitedefaultmidpunct}
{\mcitedefaultendpunct}{\mcitedefaultseppunct}\relax
\EndOfBibitem
\bibitem[Garniron \latin{et~al.}(2017)Garniron, Scemama, Loos, and
  Caffarel]{Garniron.2017.10.1063/1.4992127}
Garniron,~Y.; Scemama,~A.; Loos,~P.-F.; Caffarel,~M. Hybrid
  stochastic-deterministic calculation of the second-order perturbative
  contribution of multireference perturbation theory. \emph{J. Chem. Phys.}
  \textbf{2017}, \emph{147}, 034101, DOI: \doi{10.1063/1.4992127}\relax
\mciteBstWouldAddEndPuncttrue
\mciteSetBstMidEndSepPunct{\mcitedefaultmidpunct}
{\mcitedefaultendpunct}{\mcitedefaultseppunct}\relax
\EndOfBibitem
\bibitem[Garniron \latin{et~al.}(2019)Garniron, Applencourt, Gasperich, Benali,
  Ferte, Paquier, Pradines, Assaraf, Reinhardt, Toulouse, Barbaresco, Renon,
  David, Malrieu, Veril, Caffarel, Loos, Giner, and
  Scemama]{Garniron.2019.10.1021/acs.jctc.9b00176}
Garniron,~Y.; Applencourt,~T.; Gasperich,~K.; Benali,~A.; Ferte,~A.;
  Paquier,~J.; Pradines,~B.; Assaraf,~R.; Reinhardt,~P.; Toulouse,~J.
  \latin{et~al.}  Quantum Package 2.0: An Open-Source Determinant-Driven Suite
  of Programs. \emph{J. Chem. Theory Comput.} \textbf{2019}, \emph{15},
  3591--3609, DOI: \doi{10.1021/acs.jctc.9b00176}\relax
\mciteBstWouldAddEndPuncttrue
\mciteSetBstMidEndSepPunct{\mcitedefaultmidpunct}
{\mcitedefaultendpunct}{\mcitedefaultseppunct}\relax
\EndOfBibitem
\bibitem[Schriber and Evangelista(2016)Schriber, and
  Evangelista]{Schriber.2016.10.1063/1.4948308}
Schriber,~J.~B.; Evangelista,~F.~A. Communication: An Adaptive Configuration
  Interaction Approach for Strongly Correlated Electrons with Tunable Accuracy.
  \emph{J. Chem. Phys.} \textbf{2016}, \emph{144}, 161106, DOI:
  \doi{10.1063/1.4948308}\relax
\mciteBstWouldAddEndPuncttrue
\mciteSetBstMidEndSepPunct{\mcitedefaultmidpunct}
{\mcitedefaultendpunct}{\mcitedefaultseppunct}\relax
\EndOfBibitem
\bibitem[Schriber and Evangelista(2017)Schriber, and
  Evangelista]{Schriber.2017.10.1021/acs.jctc.7b00725}
Schriber,~J.~B.; Evangelista,~F.~A. Adaptive Configuration Interaction for
  Computing Challenging Electronic Excited States with Tunable Accuracy.
  \emph{J. Chem. Theory Comput.} \textbf{2017}, \emph{13}, 5354--5366, DOI:
  \doi{10.1021/acs.jctc.7b00725}\relax
\mciteBstWouldAddEndPuncttrue
\mciteSetBstMidEndSepPunct{\mcitedefaultmidpunct}
{\mcitedefaultendpunct}{\mcitedefaultseppunct}\relax
\EndOfBibitem
\bibitem[Tubman \latin{et~al.}(2016)Tubman, Lee, Takeshita, Head-Gordon, and
  Whaley]{Tubman.2016.10.1063/1.4955109}
Tubman,~N.~M.; Lee,~J.; Takeshita,~T.~Y.; Head-Gordon,~M.; Whaley,~K.~B. A
  Deterministic Alternative to the Full Configuration Interaction Quantum Monte
  Carlo Method. \emph{J. Chem. Phys.} \textbf{2016}, \emph{145}, 044112, DOI:
  \doi{10.1063/1.4955109}\relax
\mciteBstWouldAddEndPuncttrue
\mciteSetBstMidEndSepPunct{\mcitedefaultmidpunct}
{\mcitedefaultendpunct}{\mcitedefaultseppunct}\relax
\EndOfBibitem
\bibitem[Paldus \latin{et~al.}(1984)Paldus, {\v{C}}{\'{i}}{\v{z}}ek, and
  Takahashi]{Paldus.1984.10.1103/PhysRevA.30.2193}
Paldus,~J.; {\v{C}}{\'{i}}{\v{z}}ek,~J.; Takahashi,~M. Approximate Account of
  the Connected Quadruply Excited Clusters in the Coupled-Pair Many-Electron
  Theory. \emph{Phys. Rev. A} \textbf{1984}, \emph{30}, 2193--2209, DOI:
  \doi{10.1103/PhysRevA.30.2193}\relax
\mciteBstWouldAddEndPuncttrue
\mciteSetBstMidEndSepPunct{\mcitedefaultmidpunct}
{\mcitedefaultendpunct}{\mcitedefaultseppunct}\relax
\EndOfBibitem
\bibitem[Piecuch and Paldus(1990)Piecuch, and
  Paldus]{Piecuch.1990.10.1007/BF01119191}
Piecuch,~P.; Paldus,~J. Coupled Cluster Approaches with an Approximate Account
  of Triexcitations and the Optimized Inner Projection Technique. \emph{Theor.
  Chim. Acta} \textbf{1990}, \emph{78}, 65--128, DOI:
  \doi{10.1007/BF01119191}\relax
\mciteBstWouldAddEndPuncttrue
\mciteSetBstMidEndSepPunct{\mcitedefaultmidpunct}
{\mcitedefaultendpunct}{\mcitedefaultseppunct}\relax
\EndOfBibitem
\bibitem[Piecuch \latin{et~al.}(1996)Piecuch, Tobo\l{}a, and
  Paldus]{Piecuch.1996.10.1103/PhysRevA.54.1210}
Piecuch,~P.; Tobo\l{}a,~R.; Paldus,~J. Approximate Account of Connected
  Quadruply Excited Clusters in Single-Reference Coupled-Cluster Theory via
  Cluster Analysis of the Projected Unrestricted Hartree-Fock Wave Function.
  \emph{Phys. Rev. A} \textbf{1996}, \emph{54}, 1210--1241, DOI:
  \doi{10.1103/PhysRevA.54.1210}\relax
\mciteBstWouldAddEndPuncttrue
\mciteSetBstMidEndSepPunct{\mcitedefaultmidpunct}
{\mcitedefaultendpunct}{\mcitedefaultseppunct}\relax
\EndOfBibitem
\bibitem[Paldus and Planelles(1994)Paldus, and
  Planelles]{Paldus.1994.10.1007/BF01123868}
Paldus,~J.; Planelles,~J. Valence Bond Corrected Single Reference Coupled
  Cluster Approach. \emph{Theor. Chim. Acta} \textbf{1994}, \emph{89}, 13--31,
  DOI: \doi{10.1007/BF01123868}\relax
\mciteBstWouldAddEndPuncttrue
\mciteSetBstMidEndSepPunct{\mcitedefaultmidpunct}
{\mcitedefaultendpunct}{\mcitedefaultseppunct}\relax
\EndOfBibitem
\bibitem[Stolarczyk(1994)]{Stolarczyk.1994.10.1016/0009-2614(93)E1333-C}
Stolarczyk,~L.~Z. Complete Active Space Coupled-Cluster Method. Extension of
  Single-Reference Coupled-Cluster Method using the {CASSCF} Wavefunction.
  \emph{Chem. Phys. Lett.} \textbf{1994}, \emph{217}, 1--6, DOI:
  \doi{10.1016/0009-2614(93)E1333-C}\relax
\mciteBstWouldAddEndPuncttrue
\mciteSetBstMidEndSepPunct{\mcitedefaultmidpunct}
{\mcitedefaultendpunct}{\mcitedefaultseppunct}\relax
\EndOfBibitem
\bibitem[Peris \latin{et~al.}(1997)Peris, Planelles, and
  Paldus]{Peris.1997.10.1002/(SICI)1097-461X(1997)62:2<137::AID-QUA2>3.0.CO;2-X}
Peris,~G.; Planelles,~J.; Paldus,~J. Single-Reference {CCSD} Approach Employing
  Three- and Four-Body {CAS} {SCF} Corrections: A Preliminary Study of a Simple
  Model. \emph{Int. J. Quantum Chem.} \textbf{1997}, \emph{62}, 137--151, DOI:
  \doi{10.1002/(SICI)1097-461X(1997)62:2<137::AID-QUA2>3.0.CO;2-X}\relax
\mciteBstWouldAddEndPuncttrue
\mciteSetBstMidEndSepPunct{\mcitedefaultmidpunct}
{\mcitedefaultendpunct}{\mcitedefaultseppunct}\relax
\EndOfBibitem
\bibitem[Peris \latin{et~al.}(1999)Peris, Planelles, Malrieu, and
  Paldus]{Peris.1999.10.1063/1.479116}
Peris,~G.; Planelles,~J.; Malrieu,~J.-P.; Paldus,~J. Perturbatively Selected
  {CI} as an Optimal Source for Externally Corrected {CCSD}. \emph{J. Chem.
  Phys.} \textbf{1999}, \emph{110}, 11708--11716, DOI:
  \doi{10.1063/1.479116}\relax
\mciteBstWouldAddEndPuncttrue
\mciteSetBstMidEndSepPunct{\mcitedefaultmidpunct}
{\mcitedefaultendpunct}{\mcitedefaultseppunct}\relax
\EndOfBibitem
\bibitem[Li and Paldus(1997)Li, and Paldus]{Li.1997.10.1063/1.474289}
Li,~X.; Paldus,~J. Reduced Multireference {CCSD} Method: An Effective Approach
  to Quasidegenerate States. \emph{J. Chem. Phys.} \textbf{1997}, \emph{107},
  6257--6269, DOI: \doi{10.1063/1.474289}\relax
\mciteBstWouldAddEndPuncttrue
\mciteSetBstMidEndSepPunct{\mcitedefaultmidpunct}
{\mcitedefaultendpunct}{\mcitedefaultseppunct}\relax
\EndOfBibitem
\bibitem[Li and Paldus(2006)Li, and Paldus]{Li.2006.10.1063/1.2194543}
Li,~X.; Paldus,~J. Reduced Multireference Coupled Cluster Method with Singles
  and Doubles: Perturbative Corrections for Triples. \emph{J. Chem. Phys.}
  \textbf{2006}, \emph{124}, 174101, DOI: \doi{10.1063/1.2194543}\relax
\mciteBstWouldAddEndPuncttrue
\mciteSetBstMidEndSepPunct{\mcitedefaultmidpunct}
{\mcitedefaultendpunct}{\mcitedefaultseppunct}\relax
\EndOfBibitem
\bibitem[Paldus(2017)]{Paldus.2017.10.1007/s10910-016-0688-6}
Paldus,~J. Externally and Internally Corrected Coupled Cluster Approaches: An
  Overview. \emph{J. Math. Chem.} \textbf{2017}, \emph{55}, 477--502, DOI:
  \doi{10.1007/s10910-016-0688-6}\relax
\mciteBstWouldAddEndPuncttrue
\mciteSetBstMidEndSepPunct{\mcitedefaultmidpunct}
{\mcitedefaultendpunct}{\mcitedefaultseppunct}\relax
\EndOfBibitem
\bibitem[Deustua \latin{et~al.}(2018)Deustua, Magoulas, Shen, and
  Piecuch]{Deustua.2018.10.1063/1.5055769}
Deustua,~J.~E.; Magoulas,~I.; Shen,~J.; Piecuch,~P. Communication: Approaching
  Exact Quantum Chemistry by Cluster Analysis of Full Configuration Interaction
  Quantum Monte Carlo Wave Functions. \emph{J. Chem. Phys.} \textbf{2018},
  \emph{149}, 151101, DOI: \doi{10.1063/1.5055769}\relax
\mciteBstWouldAddEndPuncttrue
\mciteSetBstMidEndSepPunct{\mcitedefaultmidpunct}
{\mcitedefaultendpunct}{\mcitedefaultseppunct}\relax
\EndOfBibitem
\bibitem[Aroeira \latin{et~al.}(2021)Aroeira, Davis, Turney, and
  Schaefer]{Aroeira.2020.10.1021/acs.jctc.0c00888}
Aroeira,~G. J.~R.; Davis,~M.~M.; Turney,~J.~M.; Schaefer,~H.~F.,~III Coupled
  Cluster Externally Corrected by Adaptive Configuration Interaction. \emph{J.
  Chem. Theory Comput.} \textbf{2021}, \emph{17}, 182--190, DOI:
  \doi{10.1021/acs.jctc.0c00888}\relax
\mciteBstWouldAddEndPuncttrue
\mciteSetBstMidEndSepPunct{\mcitedefaultmidpunct}
{\mcitedefaultendpunct}{\mcitedefaultseppunct}\relax
\EndOfBibitem
\bibitem[Magoulas \latin{et~al.}(2021)Magoulas, Gururangan, Piecuch, Deustua,
  and Shen]{Magoulas.2021.10.1021/acs.jctc.1c00181}
Magoulas,~I.; Gururangan,~K.; Piecuch,~P.; Deustua,~J.~E.; Shen,~J. Is
  Externally Corrected Coupled Cluster Always Better Than the Underlying
  Truncated Configuration Interaction? \emph{J. Chem. Theory Comput.}
  \textbf{2021}, \emph{17}, 4006--4027, DOI:
  \doi{10.1021/acs.jctc.1c00181}\relax
\mciteBstWouldAddEndPuncttrue
\mciteSetBstMidEndSepPunct{\mcitedefaultmidpunct}
{\mcitedefaultendpunct}{\mcitedefaultseppunct}\relax
\EndOfBibitem
\bibitem[Hehre \latin{et~al.}(1969)Hehre, Stewart, and
  Pople]{Hehre.1969.10.1063/1.1672392}
Hehre,~W.~J.; Stewart,~R.~F.; Pople,~J.~A. Self-Consistent Molecular-Orbital
  Methods. {I.} {Use} of {Gaussian} Expansions of {Slater}-Type Atomic
  Orbitals. \emph{J. Chem. Phys.} \textbf{1969}, \emph{51}, 2657--2664, DOI:
  \doi{10.1063/1.1672392}\relax
\mciteBstWouldAddEndPuncttrue
\mciteSetBstMidEndSepPunct{\mcitedefaultmidpunct}
{\mcitedefaultendpunct}{\mcitedefaultseppunct}\relax
\EndOfBibitem
\bibitem[Deustua \latin{et~al.}(2017)Deustua, Shen, and
  Piecuch]{Deustua.2017.10.1103/PhysRevLett.119.223003}
Deustua,~J.~E.; Shen,~J.; Piecuch,~P. Converging High-Level Coupled-Cluster
  Energetics by Monte Carlo Sampling and Moment Expansions. \emph{Phys. Rev.
  Lett.} \textbf{2017}, \emph{119}, 223003, DOI:
  \doi{10.1103/PhysRevLett.119.223003}\relax
\mciteBstWouldAddEndPuncttrue
\mciteSetBstMidEndSepPunct{\mcitedefaultmidpunct}
{\mcitedefaultendpunct}{\mcitedefaultseppunct}\relax
\EndOfBibitem
\bibitem[Gururangan \latin{et~al.}(2021)Gururangan, Deustua, Shen, and
  Piecuch]{Gururangan.2021.10.1063/5.0064400}
Gururangan,~K.; Deustua,~J.~E.; Shen,~J.; Piecuch,~P. High-Level
  Coupled-Cluster Energetics by Merging Moment Expansions with Selected
  Configuration Interaction. \emph{J. Chem. Phys.} \textbf{2021}, \emph{155},
  174114, DOI: \doi{10.1063/5.0064400}\relax
\mciteBstWouldAddEndPuncttrue
\mciteSetBstMidEndSepPunct{\mcitedefaultmidpunct}
{\mcitedefaultendpunct}{\mcitedefaultseppunct}\relax
\EndOfBibitem
\bibitem[Pulay(1980)]{Pulay.1980.10.1016/0009-2614(80)80396-4}
Pulay,~P. Convergence Acceleration of Iterative Sequences. {The} Case of {SCF}
  Iteration. \emph{Chem. Phys. Lett.} \textbf{1980}, \emph{73}, 393--398, DOI:
  \doi{10.1016/0009-2614(80)80396-4}\relax
\mciteBstWouldAddEndPuncttrue
\mciteSetBstMidEndSepPunct{\mcitedefaultmidpunct}
{\mcitedefaultendpunct}{\mcitedefaultseppunct}\relax
\EndOfBibitem
\bibitem[Pulay(1982)]{Pulay.1982.10.1002/jcc.540030413}
Pulay,~P. Improved {SCF} Convergence Acceleration. \emph{J. Comput. Chem.}
  \textbf{1982}, \emph{3}, 556--560, DOI: \doi{10.1002/jcc.540030413}\relax
\mciteBstWouldAddEndPuncttrue
\mciteSetBstMidEndSepPunct{\mcitedefaultmidpunct}
{\mcitedefaultendpunct}{\mcitedefaultseppunct}\relax
\EndOfBibitem
\bibitem[Scuseria \latin{et~al.}(1986)Scuseria, Lee, and {Schaefer
  III}]{Scuseria.1986.10.1016/0009-2614(86)80461-4}
Scuseria,~G.~E.; Lee,~T.~J.; {Schaefer III},~H.~F. Accelerating the Convergence
  of the Coupled-Cluster Approach. \emph{Chem. Phys. Lett.} \textbf{1986},
  \emph{130}, 236--239, DOI: \doi{10.1016/0009-2614(86)80461-4}\relax
\mciteBstWouldAddEndPuncttrue
\mciteSetBstMidEndSepPunct{\mcitedefaultmidpunct}
{\mcitedefaultendpunct}{\mcitedefaultseppunct}\relax
\EndOfBibitem
\bibitem[Nooijen(2000)]{Nooijen.2000.10.1103/PhysRevLett.84.2108}
Nooijen,~M. Can the Eigenstates of a Many-Body {Hamiltonian} Be Represented
  Exactly Using a General Two-Body Cluster Expansion? \emph{Phys. Rev. Lett.}
  \textbf{2000}, \emph{84}, 2108--2111, DOI:
  \doi{10.1103/PhysRevLett.84.2108}\relax
\mciteBstWouldAddEndPuncttrue
\mciteSetBstMidEndSepPunct{\mcitedefaultmidpunct}
{\mcitedefaultendpunct}{\mcitedefaultseppunct}\relax
\EndOfBibitem
\bibitem[Nakatsuji(2000)]{Nakatsuji.2000.10.1063/1.1287275}
Nakatsuji,~H. Structure of the Exact Wave Function. \emph{J. Chem. Phys.}
  \textbf{2000}, \emph{113}, 2949--2956, DOI: \doi{10.1063/1.1287275}\relax
\mciteBstWouldAddEndPuncttrue
\mciteSetBstMidEndSepPunct{\mcitedefaultmidpunct}
{\mcitedefaultendpunct}{\mcitedefaultseppunct}\relax
\EndOfBibitem
\bibitem[Stair and Evangelista(2022)Stair, and
  Evangelista]{Stair.2022.10.1021/acs.jctc.1c01155}
Stair,~N.~H.; Evangelista,~F.~A. {QForte}: An Efficient State-Vector Emulator
  and Quantum Algorithms Library for Molecular Electronic Structure. \emph{J.
  Chem. Theory Comput.} \textbf{2022}, \emph{18}, 1555--1568, DOI:
  \doi{10.1021/acs.jctc.1c01155}\relax
\mciteBstWouldAddEndPuncttrue
\mciteSetBstMidEndSepPunct{\mcitedefaultmidpunct}
{\mcitedefaultendpunct}{\mcitedefaultseppunct}\relax
\EndOfBibitem
\bibitem[Smith \latin{et~al.}(2020)Smith, Burns, Simmonett, Parrish, Schieber,
  Galvelis, Kraus, Kruse, Remigio, Alenaizan, James, Lehtola, Misiewicz,
  Scheurer, Shaw, Schriber, Xie, Glick, Sirianni, O'Brien, Waldrop, Kumar,
  Hohenstein, Pritchard, Brooks, {Schaefer III}, Sokolov, Patkowski, {DePrince
  III}, Bozkaya, King, Evangelista, Turney, Crawford, and
  Sherrill]{Smith.2020.10.1063/5.0006002}
Smith,~D. G.~A.; Burns,~L.~A.; Simmonett,~A.~C.; Parrish,~R.~M.;
  Schieber,~M.~C.; Galvelis,~R.; Kraus,~P.; Kruse,~H.; Remigio,~R.~D.;
  Alenaizan,~A. \latin{et~al.}  PSI4 1.4: Open-Source Software for
  High-Throughput Quantum Chemistry. \emph{J. Chem. Phys.} \textbf{2020},
  \emph{152}, 184108, DOI: \doi{10.1063/5.0006002}\relax
\mciteBstWouldAddEndPuncttrue
\mciteSetBstMidEndSepPunct{\mcitedefaultmidpunct}
{\mcitedefaultendpunct}{\mcitedefaultseppunct}\relax
\EndOfBibitem
\bibitem[Kowalski and Piecuch(2004)Kowalski, and
  Piecuch]{Kowalski.2004.10.1063/1.1632474}
Kowalski,~K.; Piecuch,~P. New Coupled-Cluster Methods with Singles, Doubles,
  and Noniterative Triples for High Accuracy Calculations of Excited Electronic
  States. \emph{J. Chem. Phys.} \textbf{2004}, \emph{120}, 1715--1738, DOI:
  \doi{10.1063/1.1632474}\relax
\mciteBstWouldAddEndPuncttrue
\mciteSetBstMidEndSepPunct{\mcitedefaultmidpunct}
{\mcitedefaultendpunct}{\mcitedefaultseppunct}\relax
\EndOfBibitem
\bibitem[Lutz and Piecuch(2014)Lutz, and
  Piecuch]{Lutz.2014.10.1016/j.comptc.2014.05.008}
Lutz,~J.~J.; Piecuch,~P. Performance of the Completely Renormalized
  Equation-of-Motion Coupled-Cluster Method in Calculations of Excited-State
  Potential Cuts of Water. \emph{Comput. Theor. Chem.} \textbf{2014},
  \emph{1040--1041}, 20--34, DOI: \doi{10.1016/j.comptc.2014.05.008}\relax
\mciteBstWouldAddEndPuncttrue
\mciteSetBstMidEndSepPunct{\mcitedefaultmidpunct}
{\mcitedefaultendpunct}{\mcitedefaultseppunct}\relax
\EndOfBibitem
\bibitem[Piecuch \latin{et~al.}(2015)Piecuch, Hansen, and
  Ajala]{Piecuch.2015.10.1080/00268976.2015.1076901}
Piecuch,~P.; Hansen,~J.~A.; Ajala,~A.~O. Benchmarking the Completely
  Renormalised Equation-of-Motion Coupled-Cluster Approaches for Vertical
  Excitation Energies. \emph{Mol. Phys.} \textbf{2015}, \emph{113}, 3085--3127,
  DOI: \doi{10.1080/00268976.2015.1076901}\relax
\mciteBstWouldAddEndPuncttrue
\mciteSetBstMidEndSepPunct{\mcitedefaultmidpunct}
{\mcitedefaultendpunct}{\mcitedefaultseppunct}\relax
\EndOfBibitem
\bibitem[W{\l}och \latin{et~al.}(2005)W{\l}och, Gour, Kowalski, and
  Piecuch]{Wloch.2005.10.1063/1.1924596}
W{\l}och,~M.; Gour,~J.~R.; Kowalski,~K.; Piecuch,~P. Extension of Renormalized
  Coupled-Cluster Methods including Triple Excitations to Excited Electronic
  States of Open-Shell Molecules. \emph{J. Chem. Phys.} \textbf{2005},
  \emph{122}, 214107, DOI: \doi{10.1063/1.1924596}\relax
\mciteBstWouldAddEndPuncttrue
\mciteSetBstMidEndSepPunct{\mcitedefaultmidpunct}
{\mcitedefaultendpunct}{\mcitedefaultseppunct}\relax
\EndOfBibitem
\bibitem[Fradelos \latin{et~al.}(2011)Fradelos, Lutz, Weso{\l}owski, and
  Piecuch]{Fradelos.2011.10.1021/ct200101x}
Fradelos,~G.; Lutz,~J.~J.; Weso{\l}owski,~T.~A.; Piecuch,~P. Embedding vs
  Supermolecular Strategies in Evaluating the Hydrogen-Bonding-Induced Shifts
  of Excitation Energies. \emph{J. Chem. Theory Comput.} \textbf{2011},
  \emph{7}, 1647--1666, DOI: \doi{10.1021/ct200101x}\relax
\mciteBstWouldAddEndPuncttrue
\mciteSetBstMidEndSepPunct{\mcitedefaultmidpunct}
{\mcitedefaultendpunct}{\mcitedefaultseppunct}\relax
\EndOfBibitem
\bibitem[Yuwono \latin{et~al.}(2020)Yuwono, Chakraborty, Deustua, Shen, and
  Piecuch]{Yuwono.2020.10.1080/00268976.2020.1817592}
Yuwono,~S.~H.; Chakraborty,~A.; Deustua,~J.~E.; Shen,~J.; Piecuch,~P.
  Accelerating Convergence of Equation-of-Motion Coupled-Cluster Computations
  using the Semi-Stochastic CC($P$;$Q$) Formalism. \emph{Mol. Phys.}
  \textbf{2020}, \emph{118}, e1817592, DOI:
  \doi{10.1080/00268976.2020.1817592}\relax
\mciteBstWouldAddEndPuncttrue
\mciteSetBstMidEndSepPunct{\mcitedefaultmidpunct}
{\mcitedefaultendpunct}{\mcitedefaultseppunct}\relax
\EndOfBibitem
\end{mcitethebibliography}

\end{document}


\newpage
\noindent

This Supporting Information document is organized as follows. In Section \ref{sec_mmcc_consistency}
we provide the proof and underlying conditions for the size-consistency
of the non-iterative energy corrections, derived from the formalism of the method of moments of
coupled-cluster (MMCC) equations, considered in the main text. Section \ref{sec_additional_results} contains,
in a graphical form, additional numerical results.

The numerical data generated in this study can be found in the Excel file that forms part of the
present Supporting Information.

\section{Method of Moments of Coupled-Cluster Equations and Unitary Coupled Cluster: Size Consistency}\label{sec_mmcc_consistency}

	Here, we provide a mathematical proof of the size consistency of the MMUCC
formulas considered in the main text, along with the associated necessary and sufficient conditions.

We begin our analysis by examining the isolated systems \textit{A} and \textit{B}, characterized by the $H_A$
and $H_B$ Hamiltonians, respectively. The Schr\"{o}dinger equations for the ground electronic
states of each of the two systems read
\begin{equation} \label{SE_A}
	H_A \ket*{\Psi_{0,A}} = E_{0,A} \ket*{\Psi_{0,A}}
\end{equation}
and
\begin{equation}\label{SE_B}
	H_B \ket*{\Psi_{0,B}} = E_{0,B} \ket*{\Psi_{0,B}}.
\end{equation}
The exact $\ket*{\Psi_{0,A}}$ and $\ket*{\Psi_{0,B}}$ electronic states can be expressed
in terms of the following unitary parametrizations:
\begin{equation}\label{exact_ansatz_A}
	\ket*{\Psi_{0,A}} = U_A \ket*{\Phi_A}
\end{equation}
and
\begin{equation}\label{exact_ansatz_B}
	\ket*{\Psi_{0,B}} = U_B \ket*{\Phi_B},
\end{equation}
with $U_A$ and $U_B$ representing products of elementary unitary excitation operators, and $\ket*{\Phi_A}$
and $\ket*{\Phi_B}$ denoting the reference Slater determinants of systems
\textit{A} and \textit{B}, respectively. The corresponding sets of all excited determinants afforded by the
respective one-electron bases are denoted $\{\ket*{\Phi_{k,A}}\}$ for system \textit{A} and
$\{\ket*{\Phi_{k,B}}\}$ in the case of system \textit{B}. Without loss of generality,
we assume the use of canonical Hartree--Fock orbitals. The many-electron Hilbert space of system \textit{A},
$\mathcal{H}_A$, can be partitioned as $\mathcal{H}_A = \mathcal{H}_{0,A} \oplus \mathcal{H}_{K,A}$,
where $\mathcal{H}_{0,A} = \spn (\ket*{\Phi_A})$ and $\mathcal{H}_{K,A} = \spn (\{\ket*{\Phi_{k,A}}\})$.
Similar definitions apply in the case of system \textit{B}. 
The fact that $\ket*{\Psi_{0,A}}$ and $\ket*{\Psi_{0,B}}$ are exact implies the following
residual conditions:
\begin{equation}
	\mel*{\Phi_{k,A}}{\bar{H}_A}{\Phi_A} = 0, \forall \ket*{\Phi_{k,A}}  \in \mathcal{H}_{K,A}
\end{equation}
and
\begin{equation}
	\mel*{\Phi_{k,B}}{\bar{H}_B}{\Phi_B} = 0, \forall \ket*{\Phi_{k,B}} \in \mathcal{H}_{K,B},
\end{equation}
where $\bar{H}_A = U_A^\dagger H_A U_A$ and $\bar{H}_B = U_B^\dagger H_B U_B$ are the pertinent
similarity transformed Hamiltonians.

Subsequently, we turn our attention to an approximate description of the isolated systems \textit{A} and \textit{B}.
In this case, the $U_A^{(P_A)}$ and $U_B^{(P_B)}$ unitaries that define the $\ket*{\Psi_{0,A}^{(P_A)}}$
and $\ket*{\Psi_{0,B}^{(P_B)}}$ ans\"{a}tze,
\begin{equation}
	\ket*{\Psi_{0,A}^{(P_A)}} = U_A^{(P_A)} \ket*{\Phi_A}
\end{equation}
and
\begin{equation}
	\ket*{\Psi_{0,B}^{(P_B)}} = U_B^{(P_B)} \ket*{\Phi_B},
\end{equation}
contain fewer parameters than the dimensions of the $\mathcal{H}_{K,A}$
and $\mathcal{H}_{K,B}$ spaces, respectively. Consequently, the residual conditions can only be satisfied in the
$\mathcal{H}_{P,A} \subset \mathcal{H}_{K,A}$ and $\mathcal{H}_{P,B} \subset \mathcal{H}_{K,B}$
subspaces corresponding to the excitation operators appearing in the $U_A^{(P_A)}$ and $U_B^{(P_B)}$ unitaries:
\begin{equation}
	\mel*{\Phi_{p,A}}{\bar{H}^{(P_A)}}{\Phi_A} = 0, \forall \ket*{\Phi_{p,A}} \in \mathcal{H}_{P,A}
\end{equation}
and
\begin{equation}
	\mel*{\Phi_{p,B}}{\bar{H}^{(P_B)}}{\Phi_B} = 0, \forall \ket*{\Phi_{p,B}} \in \mathcal{H}_{P,B}.
\end{equation}
The $\mathcal{H}_{Q,A} \equiv \mathcal{H}_{K,A} \ominus \mathcal{H}_{P,A}$
and $\mathcal{H}_{Q,B} \equiv \mathcal{H}_{K,B} \ominus \mathcal{H}_{P,B}$
subspaces are spanned by the excited Slater determinants $\{\ket*{\Phi_{q,A}}\}$ and
$\{\ket*{\Phi_{q,B}}\}$, respectively, that do not satisfy the pertinent residual conditions.

Next we consider the electronic structure of the supersystem \textit{AB}, comprised of the
non-interacting systems \textit{A} and \textit{B}. Since there is no coupling between the two
systems, the Hamiltonian of supersystem \textit{AB} is
\begin{equation}
	H_{AB} = H_A \otimes \mathbf{1}_B + \mathbf{1}_A \otimes H_B \equiv H_A + H_B,
\end{equation}
with $\mathbf{1}_A$ and $\mathbf{1}_B$ denoting the identity operators on the $\mathcal{H}_A$
and $\mathcal{H}_B$ many-electron Hilbert spaces, respectively. Due to the fact that
factorized UCC ans\"{a}tze are not orbital-invariant, the choice of orbitals is crucial for achieving
size consistency. To that end, we consider a localized orbital basis, comprised of the canonical
Hartree--Fock orbitals of systems \textit{A} and \textit{B}. The use of a localized basis and the fact
that systems \textit{A} and \textit{B} do not interact imply that the unitaries defining the exact
and approximate ans\"{a}tze of supersystem \textit{AB} will only involve localized excitations, i.e.,
\begin{equation}
	\ket*{\Psi_{0,AB}} = U_{AB} \ket*{\Phi_{AB}} = U_A U_B \ket*{\Phi_{AB}}
\end{equation}
and
\begin{equation}
	\ket*{\Psi_{0,AB}^{(P_A,P_B)}} = U_{AB}^{(P_A, P_B)} \ket*{\Phi_{AB}} = U_A^{(P_A)} U_B^{(P_B)} \ket*{\Phi_{AB}},
\end{equation}
with $\ket*{\Phi_{AB}}$ being the reference determinant of the supersystem \textit{AB}. Furthermore, the operators $H_A$,
$U_A$, and $U_A^{(P_A)}$ and $H_B$, $U_B$, and $U_B^{(P_B)}$ pairwise commute since they are acting on different spaces.
Note that to ensure the proper multiplicative separability of the wavefunction and additivity of the energy, the ordering
of the elementary anti-Hermitian excitation operators defining the $U_A$, $U_B$, $U_A^{(P_A)}$, and $U_B^{(P_B)}$
unitaries needs to be preserved between the computations for the isolated systems \textit{A} and \textit{B} and the
non-interacting supersystem \textit{AB}.
Assuming that $\ket*{\Phi_{AB}}$ is separable, namely,
\begin{equation}
	\ket*{\Phi_{AB}} = \ket*{\Phi_A} \otimes \ket*{\Phi_B} \equiv \ket*{\Phi_A}\ket*{\Phi_B},
\end{equation}
we arrive at the separability of the underlying $\ket*{\Psi_{0,AB}}$ and $\ket*{\Psi_{0,AB}^{(P_A,P_B)}}$ states:
\begin{equation}
	\ket*{\Psi_{0,AB}} = U_A \ket*{\Phi_A} U_B \ket*{\Phi_B} =
	\ket*{\Psi_{0,A}} \ket*{\Psi_{0,B}}
\end{equation}
and
\begin{equation}
	\ket*{\Psi_{0,AB}^{(P_A,P_B)}} = U_A^{(P_A)} \ket*{\Phi_A} U_B^{(P_B)} \ket*{\Phi_B} =
	\ket*{\Psi_{0,A}^{(P_A)}} \ket*{\Psi_{0,B}^{(P_B)}}.
\end{equation}

First, we consider the $\delta_\text{II}$ MMCC correction, eq (16) of the main text,
since the proof is somewhat more involved.
It is straightforward to show that in the case of the non-interacting supersystem \textit{AB},
$\delta_\text{II,0}^{(P_A,P_B)}$ becomes
\begin{equation}
	\begin{split}\label{eq_full_supersystem_delta}
		\delta_{\text{II},0}^{(P_A, P_B)} =& \sum_{\ket*{\Phi_{q,A}} \in \mathcal{H}_{Q,A}}
		{\frac{\mel*{\Psi_0}{U^{(P_A, P_B)}}{\Phi_{q,A}} \ket*{\Phi_B} \bra*{\Phi_{q,A}}\bra*{\Phi_B} \bar{H}^{(P_A, P_B)}
				\ket*{\Phi_A}\ket*{\Phi_B}}{\braket*{\Psi_0}{\Psi_0^{(P_A, P_B)}}}} +\\
		&\sum_{\ket*{\Phi_{q,B}} \in \mathcal{H}_{Q,B}}
		{\frac{\mel*{\Psi_0}{U^{(P_A, P_B)}}{\Phi_A} \ket*{\Phi_{q,B}} \bra*{\Phi_A}\bra*{\Phi_{q,B}} \bar{H}^{(P_A, P_B)}
				\ket*{\Phi_A}\ket*{\Phi_B}}{\braket*{\Psi_0}{\Psi_0^{(P_A, P_B)}}}} +\\
		&\sum_{\substack{\ket*{\Phi_{q,A}} \in \mathcal{H}_{Q,A} \\ \ket*{\Phi_{q,B}} \in \mathcal{H}_{Q,B}}}
		{\frac{\mel*{\Psi_0}{U^{(P_A, P_B)}}{\Phi_{q,A}} \ket*{\Phi_{q,B}} \bra*{\Phi_{q,A}}\bra*{\Phi_{q,B}} \bar{H}^{(P_A, P_B)}
				\ket*{\Phi_A}\ket*{\Phi_B}}{\braket*{\Psi_0}{\Psi_0^{(P_A, P_B)}}}}
	\end{split}.
\end{equation}
The above expression can be simplified since
\begin{equation}
	\begin{split}
		\bra*{\Phi_{q,A}}\bra*{\Phi_{q,B}} \bar{H}^{(P_A, P_B)} \ket*{\Phi_A}\ket*{\Phi_B} &=
		\bra*{\Phi_{q,A}}\bra*{\Phi_{q,B}}(\bar{H}_A^{(P_A)} + \bar{H}_B^{(P_B)})\ket*{\Phi_A}\ket*{\Phi_B} \\
		&= \mel{\Phi_{q,A}}{\bar{H}_A^{(P_A)}}{\Phi_A}\braket{\Phi_{q,B}}{\Phi_B} +
		\mel{\Phi_{q,B}}{\bar{H}_B^{(P_B)}}{\Phi_B}\braket{\Phi_{q,A}}{\Phi_A} \\
		&= 0,
	\end{split}
\end{equation}
where in the last step we used the fact that Slater determinants are orthonormal. Therefore, eq
\eqref{eq_full_supersystem_delta} reduces to
\begin{equation}
	\begin{split}\label{eq_reduced_supersystem_delta}
		\delta_{\text{II},0}^{(P_A, P_B)} =& \sum_{\ket*{\Phi_{q,A}} \in \mathcal{H}_{Q,A}}
		{\frac{\mel*{\Psi_0}{U^{(P_A, P_B)}}{\Phi_{q,A}} \ket*{\Phi_B} \bra*{\Phi_{q,A}}\bra*{\Phi_B} \bar{H}^{(P_A, P_B)}
		\ket*{\Phi_A}\ket*{\Phi_B}}{\braket*{\Psi_0}{\Psi_0^{(P_A, P_B)}}}} +\\
		&\sum_{\ket*{\Phi_{q,B}} \in \mathcal{H}_{Q,B}}
		{\frac{\mel*{\Psi_0}{U^{(P_A, P_B)}}{\Phi_A} \ket*{\Phi_{q,B}} \bra*{\Phi_A}\bra*{\Phi_{q,B}} \bar{H}^{(P_A, P_B)}
		\ket*{\Phi_A}\ket*{\Phi_B}}{\braket*{\Psi_0}{\Psi_0^{(P_A, P_B)}}}}
	\end{split}.
\end{equation}
Before we are able to proceed any further, we need to evaluate the various quantities appearing in eq
\eqref{eq_reduced_supersystem_delta}. Starting with the denominator that is common in both terms,
it is fairly straightforward to show that it factorizes as follows:
\begin{equation}
	\braket*{\Psi_0}{\Psi_0^{(P_A, P_B)}} = \braket*{\Psi_{0,A}}{\Psi_{0,A}^{(P_A)}} \braket*{\Psi_{0,B}}{\Psi_{0,B}^{(P_B)}}.
\end{equation}
Focusing on the first term, we have that
\begin{equation}
	\begin{split}
		\mel*{\Psi_0}{U^{(P_A,P_B)}}{\Phi_{q,A}} \ket*{\Phi_B} &=
		\bra*{\Phi_A}\bra*{\Phi_B} U_A^\dagger U_B^\dagger U_A^{(P_A)} U_B^{(P_B)} \ket*{\Phi_{q,A}}\ket*{\Phi_B} \\
		&= \mel*{\Phi_A}{U_A^\dagger U_A^{(P_A)}}{\Phi_{q,A}} \mel*{\Phi_B}{U_B^\dagger U_B^{(P_B)}}{\Phi_B} \\
		&= \mel{\Psi_{0,A}}{U_A^{(P_A)}}{\Phi_{q,A}} \braket*{\Psi_{0,B}}{\Psi_{0,B}^{(P_B)}}
	\end{split}
\end{equation}
and
\begin{equation}
	\begin{split}
		\bra*{\Phi_{q,A}}\bra*{\Phi_B} \bar{H}^{(P_A, P_B)} \ket*{\Phi_A}\ket*{\Phi_B} &=
		\bra*{\Phi_{q,A}}\bra*{\Phi_B} (\bar{H_A}^{(P_A)} + \bar{H_B}^{(P_B)}) \ket*{\Phi_A}\ket*{\Phi_B} \\
		&= \mel{\Phi_{q,A}}{\bar{H}_A^{(P_A)}}{\Phi_A}\braket{\Phi_B}{\Phi_B} +
		\mel{\Phi_B}{\bar{H}_B^{(P_B)}}{\Phi_B}\braket{\Phi_{q,A}}{\Phi_A} \\
		&= \mel{\Phi_{q,A}}{\bar{H}_A^{(P_A)}}{\Phi_A}.
	\end{split}
\end{equation}
Similar expressions can be derived in the case of the second term appearing in eq \eqref{eq_reduced_supersystem_delta}.

We are now in a position to derive the final expression for the non-iterative moment correction
for the supersystem. Using the above information, eq \eqref{eq_reduced_supersystem_delta} yields
\begin{equation}
	\begin{split}
		\delta_{\text{II},0}^{(P_A, P_B)} =& \sum_{\ket*{\Phi_{q,A}} \in \mathcal{H}_{Q,A}} {\frac{\mel*{\Psi_{0,A}}{U_A^{(P_A)}}{\Phi_{q,A}}
				\braket*{\Psi_{0,B}}{\Psi_{0,B}^{(P_B)}}\mel*{\Phi_{q,A}}{\bar{H}_A^{(P_A)}}{\Phi_A}}
			{\braket*{\Psi_{0,A}}{\Psi_{0,A}^{(P_A)}}\braket*{\Psi_{0,B}}{\Psi_{0,B}^{(P_B)}}}} + \\
		&\sum_{\ket*{\Phi_{q,B}} \in \mathcal{H}_{Q,B}} {\frac{\mel*{\Psi_{0,B}}{U_B^{(P_B)}}{\Phi_{q,B}}
				\braket*{\Psi_{0,A}}{\Psi_{0,A}^{(P_A)}}\mel*{\Phi_{q,B}}{\bar{H}_B^{(P_B)}}{\Phi_B}}
			{\braket*{\Psi_{0,A}}{\Psi_{0,A}^{(P_A)}}\braket*{\Psi_{0,B}}{\Psi_{0,B}^{(P_B)}}}} \\
		=& \sum_{\ket*{\Phi_{q,A}} \in \mathcal{H}_{Q,A}} {\frac{\mel*{\Psi_{0,A}}{U_A^{(P_A)}}{\Phi_{q,A}}
				\mel*{\Phi_{q,A}}{\bar{H}_A^{(P_A)}}{\Phi_A}}
			{\braket*{\Psi_{0,A}}{\Psi_{0,A}^{(P_A)}}}} + \\
		&\sum_{\ket*{\Phi_{q,B}} \in \mathcal{H}_{Q,B}} {\frac{\mel*{\Psi_{0,B}}{U_B^{(P_B)}}{\Phi_{q,B}}
				\mel*{\Phi_{q,B}}{\bar{H}_B^{(P_B)}}{\Phi_B}}
			{\braket*{\Psi_{0,B}}{\Psi_{0,B}^{(P_B)}}}} \\
		=& \delta_{\text{II},0}^{(P_A)} + \delta_{\text{II},0}^{(P_B)}.
	\end{split}
\end{equation}

Therefore, we proved that the MMCC-type correction defined by eq (16) of the main text is size-consistent,
provided that the following conditions are satisfied:
\begin{enumerate}
	\item Use of orbitals localized on the individual fragments.
	\item Separability of the underlying reference state.
	\item The ordering of the elementary unitary excitation operators needs to be identical in the
	isolated fragments and the supersystem.
\end{enumerate}
The proof that the $\delta_\text{Ia}$ and $\delta_\text{Ic}$ non-iterative MMCC-type corrections,
given by eqs (9) and (15), respectively, of the main text, are also size-consistent follows the same steps
as above, depending on the same conditions as well.

\pagebreak

\section{Additional Numerical Results}\label{sec_additional_results}

\begin{figure*}[!h]
	\centering
	\includegraphics[width=\textwidth]{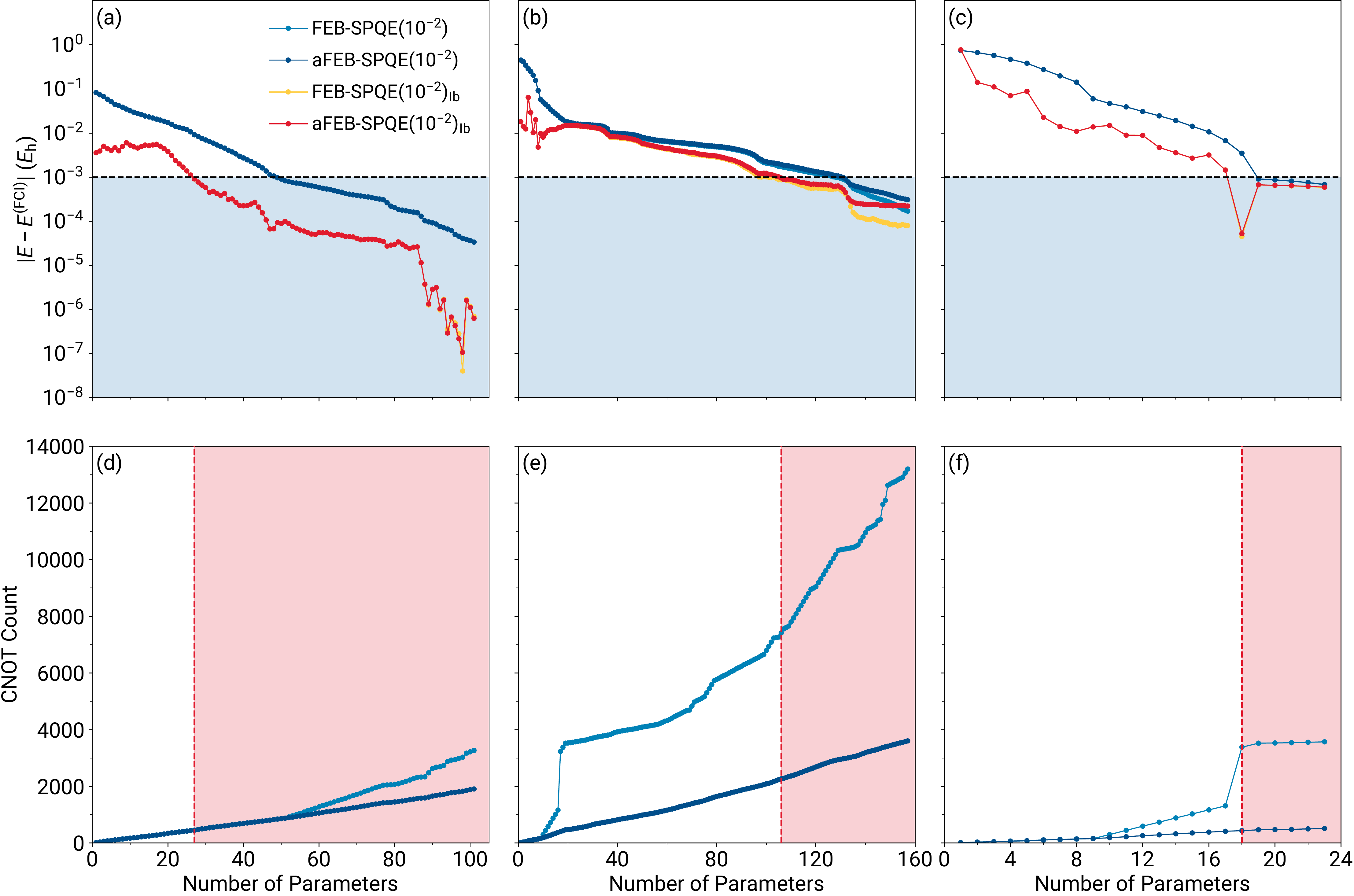}
	\caption{\label{figureS1}
		Errors relative to FCI [(a)--(c)] and CNOT gate counts [(d)--(f)] characterizing the aFEB-SPQE
		simulations of the symmetric dissociation of the $\text{H}_6$/STO-6G linear chain at three
		representative distances between neighboring H atoms, including $R_\text{H--H} = \SI{1.0}{\AA}$
		[(a) and (d)], $R_\text{H--H} = \SI{2.0}{\AA}$ [(b) and (e)], and $R_\text{H--H} = \SI{3.0}{\AA}$
		[(c) and (f)]. The blue-shaded area in the top-row panels indicates results within
		chemical accuracy (\SI{1}{\milli \textit{E}_h}) from FCI. The red-shaded area in the
		bottom-row panels denotes the CNOT counts of the underlying aFEB-SPQE quantum circuits for which
		the aFEB-SPQE$_\text{Ib}$ energetics are within chemical accuracy. To facilitate comparisons,
		the corresponding FEB-SPQE results are also included.
	}
\end{figure*}

\begin{figure*}[!h]
	\centering
	\includegraphics[width=\textwidth]{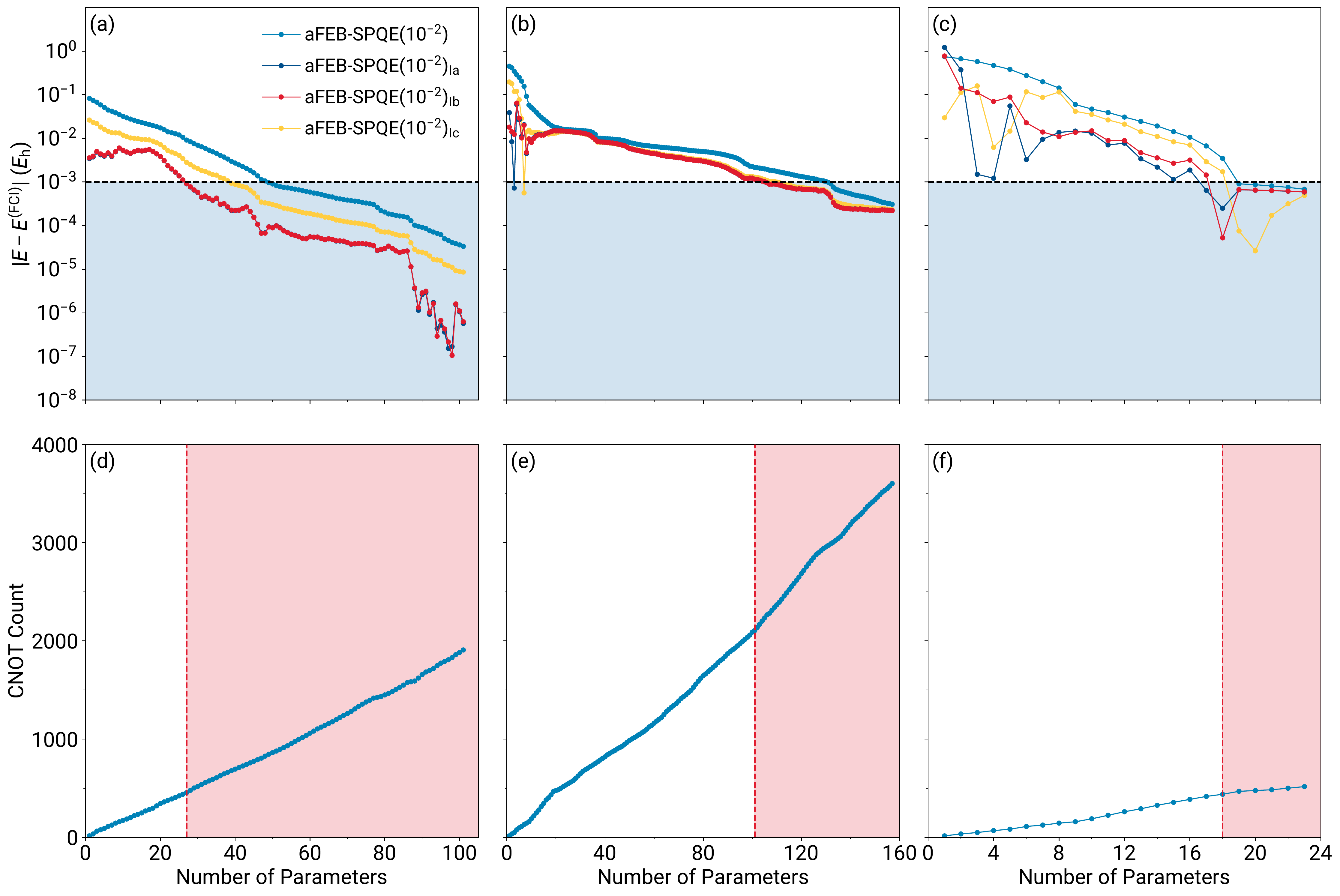}
	\caption{\label{figureS2}
		Errors relative to FCI [(a)--(c)] and CNOT gate counts [(d)--(f)] characterizing the aFEB-SPQE
		simulations of the symmetric dissociation of the $\text{H}_6$/STO-6G linear chain at three
		representative distances between neighboring H atoms, including $R_\text{H--H} = \SI{1.0}{\AA}$
		[(a) and (d)], $R_\text{H--H} = \SI{2.0}{\AA}$ [(b) and (e)], and $R_\text{H--H} = \SI{3.0}{\AA}$
		[(c) and (f)]. The blue-shaded area in the top-row panels indicates results within
		chemical accuracy (\SI{1}{\milli \textit{E}_h}) from FCI. The red-shaded area in the
		bottom-row panels denotes the CNOT counts of the underlying aFEB-SPQE quantum circuits for which
		the aFEB-SPQE$_\text{Ib}$ energies are within chemical accuracy.
	}
\end{figure*}

	\begin{figure*}[!h]
	\centering
	\includegraphics[width=\textwidth]{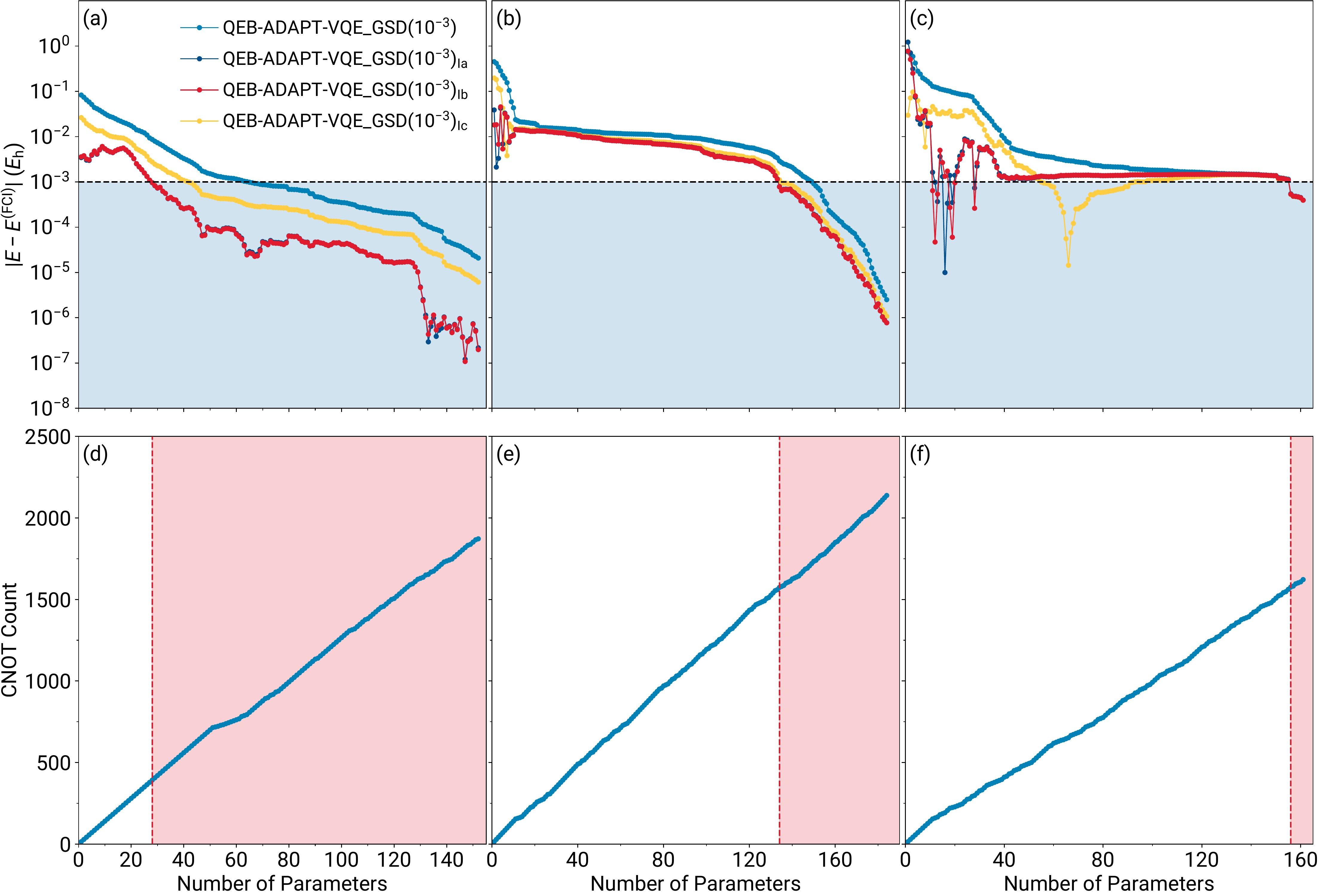}
	\caption{\label{figureS3}
		Errors relative to FCI [(a)--(c)] and CNOT gate counts [(d)--(f)] characterizing the QEB-ADAPT-VQE
		simulations of the symmetric dissociation of the $\text{H}_6$/STO-6G linear chain at three
		representative distances between neighboring H atoms, including $R_\text{H--H} = \SI{1.0}{\AA}$
		[(a) and (d)], $R_\text{H--H} = \SI{2.0}{\AA}$ [(b) and (e)], and $R_\text{H--H} = \SI{3.0}{\AA}$
		[(c) and (f)]. The blue-shaded area in the top-row panels indicates results within
		chemical accuracy (\SI{1}{\milli \textit{E}_h}) from FCI. The red-shaded area in the
		bottom-row panels denotes the CNOT counts of the underlying QEB-ADAPT-VQE quantum circuits for which
		the QEB-ADAPT-VQE$_\text{Ib}$ energies are within chemical accuracy.
	}
\end{figure*}